\documentclass[sigconf,9pt]{acmart}

\usepackage{amsmath}
\usepackage{graphicx}
\usepackage{xspace}
\usepackage{indentfirst}
\usepackage{algorithm}
\usepackage{algpseudocode}
\usepackage{multicol}
\usepackage{caption}
\usepackage{makecell}
\usepackage{subcaption}
\usepackage{xspace}
\usepackage{enumitem}
\usepackage{mdframed}
\usepackage{mathtools}
\usepackage{float}
\usepackage[small,compact]{titlesec}

\renewcommand\footnotetextcopyrightpermission[1]{} 
\setcopyright{none}

\acmPrice{}

\newcommand{\sysname}{Opus\xspace}

\providecommand{\ie}{\emph{i.e.,} }
\providecommand{\eg}{\emph{e.g.,} }

\providecommand{\etc}{\emph{etc.}}      
\providecommand{\myparab}[1]{\noindent\textbf{#1} }

\newcommand{\rev}[1]{#1}
\newcommand{\del}[1]{}
\providecommand{\myparab}[1]{\smallskip\noindent\textbf{#1} }
\providecommand{\alltoall}{\textsc{AllToAll}\xspace}
\providecommand{\allreduce}{\textsc{AllReduce}\xspace}
\providecommand{\allgather}{\textsc{AllGather}\xspace}

\providecommand{\sendrecv}{\textsc{Send/Recv}\xspace}
\providecommand{\reducescatter}{\textsc{ReduceScatter}\xspace}
\newlist{compactitem}{itemize}{1}
\setlist[compactitem,1]{label=\textbullet, left=0pt, itemsep=1pt, topsep=1pt, parsep=0pt, partopsep=0pt}

\expandafter\def\expandafter\normalsize\expandafter{%
    \normalsize%
    \setlength\abovedisplayskip{1pt}%
    \setlength\belowdisplayskip{2pt}%
    \setlength\abovedisplayshortskip{-2pt}%
    \setlength\belowdisplayshortskip{2pt}%
}

\begin{document}

\acmYear{2026}\copyrightyear{2026}
\acmConference[]{}
\acmBooktitle{}
\acmDOI{}
\acmISBN{}

\fancypagestyle{standardpagestyle}{%
  \fancyhf{}%
  \fancyfoot[C]{\thepage}%
  \renewcommand{\headrulewidth}{0pt}%
}
\fancypagestyle{firstpagestyle}{%
  \fancyhf{}%
  \fancyfoot[C]{\thepage}%
  \renewcommand{\headrulewidth}{0pt}%
}
\pagestyle{standardpagestyle}

\title{\sysname: Photonic Rail-Optimized Fabric in ML Datacenters}


\author{Eric Ding}
\affiliation{%
  \institution{Cornell University}%
  \city{Ithaca, NY}
  \country{USA}%
}

\author{Barry Lyu}
\affiliation{%
  \institution{University of Michigan}%
  \city{Ann Arbor, MI}
  \country{USA}%
}

\author{Bhaskar Kataria}
\affiliation{%
  \institution{Cornell University}%
  \city{Ithaca, NY}
  \country{USA}%
}

\author{Rachee Singh}
\affiliation{%
  \institution{Cornell University}%
  \city{Ithaca, NY}
  \country{USA}%
}

\begin{abstract}
    Rail-optimized network fabrics have become the de facto datacenter scale-out fabric for large-scale ML training. However, the use of high-radix electrical switches to provide all-to-all connectivity in rails imposes substantial power and cost. We propose a rethinking of the rail abstraction by retaining its communication semantics, but realizing it using optical circuit switches. The key challenge is that optical switches support one-to-one connectivity at a time, limiting the fan-out of traffic in ML workloads using hybrid parallelisms. We overcome this through \emph{parallelism-driven rail reconfiguration}, which exploits the non-overlapping communication phases of different parallelism dimensions. This time-multiplexes a single set of physical ports across circuit configurations tailored to each phase within a training iteration. We design and implement \sysname, a control plane that orchestrates this in-job reconfiguration of photonic rails at parallelism phase boundaries, and evaluate it on a physical OCS testbed, the Perlmutter supercomputer, and in simulation at up to 2,048 GPUs. Our results show that photonic rails can achieve over $23\times$ network power reduction and $4\times$ cost savings while incurring only modest training overhead at production-relevant OCS reconfiguration latencies.
\end{abstract}

\ccsdesc[500]{Networks~Network architectures}
\ccsdesc[500]{Networks~Data center networks}
\ccsdesc[500]{Computing methodologies~Machine learning}

\keywords{Reconfigurable Networks, Optical Circuit Switching, Large Language Model, Collective Communication}

\maketitle

\section{Introduction}

\begin{figure}[h!]
    \centering
    \includegraphics[width=1\linewidth]{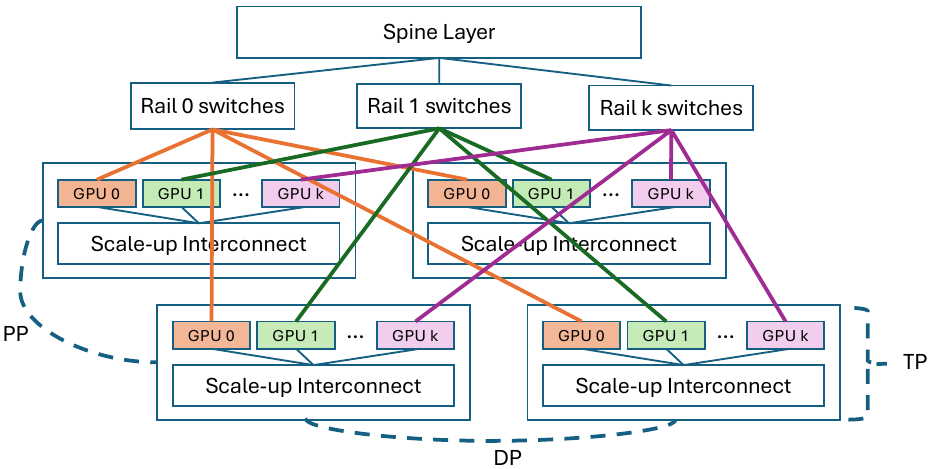}
    \caption{Rail-optimized fabrics. We propose to replace packet switches (shown as Rail 0, Rail 1 \etc) with optical circuit switches. We make the case for retaining the \emph{illusion of full connectivity} between GPU ranks connected to the same optical rail switch using in-job reconfiguration.}
    \label{fig:rail}
\end{figure}

The design of datacenter interconnect fabrics has a significant impact on the scalability and efficiency of large-scale machine learning (ML) systems. A wide range of general datacenter fabric designs have been proposed in the past decade, including electrical~\cite{vl2, jupiter-rising, harsh2020spineless,agarwal2024harmony, singla2014high} and photonic networks~\cite{farrington2010helios,cthrough,firefly,projector,chen2017enabling}. More recently, the rapid growth of ML training workloads has driven a shift toward ML-centric datacenter fabric designs~\cite{jouppi2023tpu, tpuresilience, sipac, wang2023topoopt, lumorph, morphlux, sipml}. Among the many proposals, one datacenter topology has seen broad adoption: the rail-optimized fabric~\cite{nvidia-rail-optimize, gherghescu2024ve}. The rail fabric explicitly aligns with the communication patterns of hybrid parallelisms in distributed ML by wiring together ``rails'' of GPUs---sets of GPUs with identical ranks across multiple high-bandwidth (or \emph{scale-up}) domains---to the same switch (Figure~\ref{fig:rail}). This design can achieve congestion-free communication for common collective operations like \allreduce and \allgather in distributed ML pipelines~\cite{nvidia-rail-optimize}.

But this performance comes at a steep cost. Each rail switch connects GPUs of the same rank in all scale-up domains, resulting in networks built from high-radix packet switches. These switches are not only costly, but also power-hungry. Switch ASIC processing and transceiver electrical-optical conversions contribute to the energy and complexity burden of the fabric~\cite{broadcom_scip_2021, nvidia_spectrum-x_2025, jupiter-evolving}. As ML clusters scale to tens of thousands of GPUs, the networking fabric's power footprint grows proportionally, consuming an increasingly significant share of the datacenter's total energy budget. Reducing the power consumed by the network is therefore not just a cost optimization, it is becoming an operational necessity for sustainably scaling ML infrastructure.

This paper asks whether it is possible to retain the desirable properties of rail-optimized fabrics while improving their energy and cost efficiency. Rather than redesigning the datacenter topology~\cite{wang2023topoopt, lumorph, morphlux, liao2025mixnet, slimfly,jellyfish}, we pursue a different direction to answer the question: we propose to replace electrical packet switches in the rail with reconfigurable optical circuit switches (OCSes) which consume orders of magnitude less power than their electrical counterparts~\cite{sipac, neye}. We call the resulting design a \emph{photonic rail-optimized fabric}. Our proposal draws inspiration from recent successes in optical ML fabrics~\cite{jouppi2023tpu, tpuresilience} but departs from them by preserving the widely adopted rail design, while fundamentally changing how data is switched within it.

However, the shift from packet switching to circuit switching in rails is not straightforward to achieve in practice. Electrical rail switches enable \emph{all-to-all} packet-level connectivity among GPUs in the same rail, whereas OCSes only offer one-to-one circuit connectivity, \ie a matching between GPUs, at a given time. This breaks a key invariant of rail-optimized designs---all-to-all connectivity among the same-rank GPUs across all scale-up domains (Figure~\ref{fig:rail}). Without this property, common collective operations used in hybrid parallelisms can no longer be efficiently implemented. The core technical challenge, then, is: \emph{how to retain the abstraction of rail communication despite the limitations of photonic switching?}

\sysname's key observation is that most communication in distributed ML is predictable and structured \cite{gangidi2024rdma, wang2023topoopt}. Collectives are issued in known sequences, organized by parallelism type (\eg tensor, data, pipeline)~\cite{nvidia_nccl}. We exploit this structure to break the illusion of requiring full connectivity by \emph{reconfiguring the photonic fabric between collectives within the job}. To our knowledge, this is the first proposal to reconfigure fabrics between ML collectives, while other proposals reconfigure once prior to the job start~\cite{wang2023topoopt, tpuresilience, jouppi2023tpu}, or reconfigure during workloads specifically for one type of parallelism~\cite{liao2025mixnet}.



To realize the vision of photonic rails, we develop \sysname, a control plane that orchestrates parallelism-driven rail reconfiguration. \sysname introduces a control layer between ML training frameworks and collective communication libraries. Collective communication libraries act as clients to \sysname and issue provisional intents to communicate, while \sysname interfaces with network orchestrators to reconfigure the optical fabric at parallelism phase boundaries. \sysname's design ensures that reconfiguration is both safe---circuits are never torn down while traffic is in flight---and efficient---speculative provisioning hides reconfiguration latency within natural idle windows between parallelism phases.

\sysname enables a principled quantification of the tradeoff at the heart of photonic rails: \emph{how much ML training performance must be sacrificed to achieve the power and cost savings of photonic rails?} Through experiments on a physical OCS hardware testbed, emulation on the Perlmutter supercomputer (up to 64 GPUs), and large-scale simulations (up to 2,048 GPUs), we show that this tradeoff is remarkably favorable. At production-relevant OCS reconfiguration latencies ($\leq100$ ms), \sysname incurs less than 6.7\% increase in iteration time while achieving up to $4.27\times$ reduction in networking infrastructure cost and over $23\times$ reduction in network power consumption compared to electrical rail-optimized fabrics. This shows that the performance penalty of in-job reconfiguration is a modest price for an order-of-magnitude improvement in the energy efficiency of ML datacenter networks.

\section{Our Proposal: Electrical Rails $\rightarrow$ Optical Rails}

\begin{figure}[t]
    \centering
    \includegraphics[width=1\linewidth]{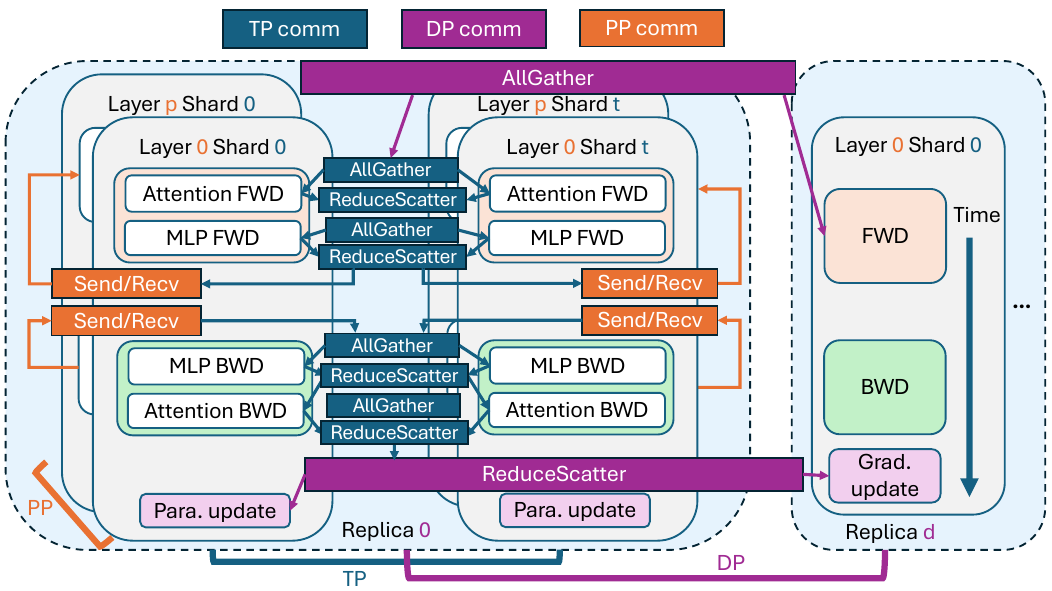}
    \caption{Traffic in a training iteration with 3D parallelism.}
    \label{fig:traffic}
    \vspace{-0.2cm}
\end{figure}

Training ML models at scale requires a combination of parallelism strategies across thousands of GPUs~\cite{chu2025scaling, rasley2020deepspeed, liu2024deepseek,shoeybi2019megatron}. To make this feasible, ML systems leverage multiple, co-existing parallelisms (Table~\ref{tab:paralllelism} in Appendix~\S\ref{sec:charac}). These include data parallelism (DP, and \del{variants like} fully sharded data parallelism or FSDP), pipeline parallelism (PP), tensor parallelism (TP, often with sequence parallelism or SP), context parallelism (CP), and expert parallelism (EP) \cite{zhao2023pytorch, rasley2020deepspeed, shoeybi2019megatron, korthikanti2023reducing, liu2023ring, fedus2022switch, liu2024deepseek, gherghescu2024ve}.  Each parallelism incurs communication that differs in: (1) \textbf{data volume:} ranging from full model size in DP to per-layer activations in TP; (2) \textbf{start time:} some collectives occur in the forward pass, others only in backpropagation; (3) \textbf{frequency:} some fire once per layer, others once per micro-batch; and (4) \textbf{communication pattern:} from symmetrical collectives, where every rank participates, to peer-to-peer asymmetrical \sendrecv (Appendix~\S\ref{sec:charac}). Importantly, the communication operations from different parallelisms are not ordered arbitrarily: they follow strict dependencies defined by the model's compute graph. Figure~\ref{fig:traffic} shows these dependencies in a 3D-parallel training step.




\rev{\myparab{Rail-optimized topologies} organize the scale-out network} into multiple independent \emph{rails}\rev{~\cite{nvidia-rail-optimize}}. Each rail connects GPUs of the same local rank across all scale-up domains (\eg an NVIDIA DGX~\cite{nvidia_dgx_superpod_2025} or
HGX~\cite{nvidia_hgx_platform_2025} node), providing a dedicated, congestion-free path for collective communication \del{that occurs}on the scale-out network, like DP and PP communication~\cite{nvidia-rail-optimize}.

Figure~\ref{fig:rail} illustrates how a 3D parallelism strategy (\ie DP, TP and PP) maps onto a rail fabric. Frequent and latency-sensitive TP collectives are confined within the high-bandwidth scale-up domains, whereas PP and DP collectives traverse the slower scale-out network since they occur less frequently and can be overlapped with compute~\cite{qi2023zero,gherghescu2024ve}. The rail abstraction allows these scale-out collectives to occur without oversubscribing links in the datacenter fabric since every rail provides a dedicated, congestion-free path among GPUs of the same rank across domains.

\myparab{Limitations of packet-switched rails.} Despite their performance benefits, today’s rail fabrics suffer from scalability challenges. Optical fibers connect scale-up domains to the rail switches. These links terminate at transceivers that interface with server NICs and packet switch ports. Each packet switch introduces optical-electrical-optical (OEO) conversions, adding energy and latency overhead to the data path. These conversions, coupled with the switch ASIC's work---packet queueing, header parsing, and TCAM lookups---consume significant energy \cite{Yeluri2023_power_consumption, semianalysis2025xaicolossus}. Moreover, while link speeds (\eg 400 Gbps) continue to scale, ASIC processing speed has not kept pace. As a result, packet-switched fabrics now represent a bottleneck in both power efficiency and bandwidth scalability, especially in the face of LLM training jobs that require hundreds of rails and thousands of GPUs~\cite{broadcom_cpo}.

\myparab{Replacing rail packet switches with OCSes.} We re-architect the rail-based datacenter fabric to replace electrical packet switches with reconfigurable optical circuit switches (OCSes). These optical switches can form end-to-end circuits without OEO conversions and eliminate switch ASICs entirely from the datapath. The resulting architecture improves on the traditional rail fabric by reducing energy consumption of the network, reducing datapath latency, and scaling bandwidth without incurring ASIC bottlenecks~\cite{Lumentum2025OCS}. Importantly, our proposal retains the physical structure of the existing rail-optimized topology: the scale-up domains, cabling, and GPU-to-rail mapping, all remain unchanged. There is no multi-tier electrical rail or spine. Instead, each rail becomes a flat, photonic point-to-point fabric. Cross-rank communication can still be supported via forwarding through the high-bandwidth interconnect in scale-up (\eg PXN~\cite{nvidia-pxn,wang2024rail})\del{, as explored in prior work~\cite{wang2024rail}}. The control plane remains electrical and host-driven.

\myparab{State-of-the-art OCS technology.}
Mature switch technologies like MEMS-based OCSes already have characteristics required by our proposed design, including millisecond reconfiguration and switch radix in the hundreds~\cite{Lumentum2025OCS, polatis_series7000, liang2024negotiator, liao2025mixnet, ballani2020sirius}. Note that our proposal differs from recent silicon photonic architectures proposed by NVIDIA and Broadcom~\cite{nvidia_silicon_photonics, broadcom_bcm78909}, which still rely on electrical switching ASICs but use co-packaged optics (CPO) instead of pluggable optical transceivers.


\section{Optical Rails: Challenges and Opportunities}
\label{sec:challenges}
Replacing electrical rail switches with OCSes is non-trivial. In ML jobs with hybrid parallelisms, each GPU participates in multiple \emph{communication groups}, \ie logical constructs managed by collective communication libraries like NCCL~\cite{nvidia_nccl_communicators}, where each group is associated with a different parallelism axis. This results in a high communication degree per GPU. For example, in a 3D-parallel job using ring-based collective communication (\eg \allreduce), each GPU requires two neighbors per ring across three parallelism dimensions, yielding a minimum degree of six. Electrical rails support high degrees of communication naturally when a GPU's dedicated NIC connects to an electrical packet switch, enabling all-to-all connectivity between connected GPUs, at all times. 

\subsection{Challenges}
In contrast, an OCS can establish only as many simultaneous circuits as a GPU has physical NICs, which is far fewer than the communication degree that hybrid parallelisms demand. Thus, photonic rails face three key challenges:
\textbf{C1: } Low degree restricts collectives to using only ring algorithms that are bandwidth-efficient but incur high latency~\cite{taccl}. Latency-optimized strategies like tree-based or recursive-doubling collectives cannot be used~\cite{thakur2003improving, sanders2009two, taccl};
\textbf{C2: } The number of parallelisms is constrained as some parallelisms (\eg CP) need dedicated connectivity~\cite{liang2024torchtitan};
\textbf{C3: } Distributing a GPU's NIC ports across communication groups allocates only a fraction of NIC bandwidth to each collective, underutilizing the provisioned bandwidth.

\myparab{A concrete example.}
Consider training a model with 3D parallelism (TP, DP, and PP) on DGX H200 nodes (scale-up) connected by an optical rail (scale-out). Each GPU has dedicated access to a ConnectX-7 NIC, which can operate in three configurations: \del{one logical}\rev{1$\times$}400 Gbps port, \del{two logical}\rev{\rev{2$\times$}}200 Gbps ports, or \del{four}\rev{4$\times$}100 Gbps ports~\cite{nvidia2025dgx_h200, nvidia_connectx7_datasheet}. \del{As is common, TP is confined to the scale-up domain. As a result, TP traffic stays within the node while DP and PP collectives traverse the scale-out optical rail.}
\rev{TP is confined to the scale-up domain, while DP and PP collectives traverse the scale-out optical rail.} 
For the DP collectives, each GPU needs to connect to two neighbors in a ring for ring-based \allreduce. For PP collectives, each GPU needs to connect to two neighbors for sending and receiving activations. Connections for both of these \del{parallelism dimensions}\rev{parallelisms} will be made between the GPU's NIC ports and the rail OCS. If we use the 4-port NIC configuration, we can assign two ports each to DP and PP to have sufficient degree for ring-based collectives to relevant neighbors. This halves the bandwidth available to collectives of each \del{parallelism dimension}\rev{parallelism}~(C3). Moreover, the low per-group degree forces the use of ring-based algorithms~(C1), and adding a fourth scale-out parallelism such as CP is infeasible without additional NICs~(C2). The root cause of these limitations is that static circuit-based connectivity of the OCS forces all \del{parallelism dimensions}\rev{parallelisms} to share a fixed set of physical ports simultaneously.

\myparab{A note on prior work.}
One workaround to these limitations is to multiplex parallelisms over shared physical links across scale-up domains. But this introduces a new set of problems: forwarding traffic via intermediate GPUs to reach the destination inflates latency and incurs a bandwidth tax~\cite{mellette2017rotornet}. Prior OCS-based ML fabrics have sidestepped these constraints by adding many more NICs per GPU~\cite{wang2023topoopt, sipac,wu2025actina}, restricting the OCS fabric to a single parallelism while relying on electrical switches for the rest~\cite{liao2025mixnet}, or constructing a fixed torus before the job starts~\cite{jouppi2023tpu}, which still suffers from C1 and C3~\cite{lumorph, morphlux}.

\subsection{Opportunities}
\label{sec:trace}
We observe that, although a GPU belongs to many communication groups, it does not use them all at once. Collectives from different \del{parallelism dimensions}\rev{parallelisms} follow strict
data dependencies defined by the model's computational graph (Figure~\ref{fig:traffic}): DP collectives occur after backpropagation completes, PP Send/Recv operations interleave
with forward and backward passes, and so on. These dependencies create brief \emph{windows} of communication inactivity between parallelism phases.

If the OCS can be reconfigured within such a window, a single set of physical ports can be time-multiplexed
across all parallelism dimensions---each phase receiving the \emph{full} NIC bandwidth and a circuit topology tailored to its collective, eliminating C1--C3 entirely. This motivates a departure from prior reconfigurable network designs, which target microsecond- or nanosecond-scale adaptation for general datacenter traffic~\cite{cthrough, projector, farrington2010helios, amir2024shale, shoal, mellette2017rotornet, liang2024negotiator,
ballani2020sirius} and are ill suited to the repetitive, high-volume collective patterns of ML workloads~\cite{wang2023topoopt}. We instead propose to reconfigure at the granularity of ML collectives themselves---a coarser but far more natural unit of adaptation for distributed training.







\begin{figure}[t]
    \centering
    \includegraphics[width=0.9\linewidth]{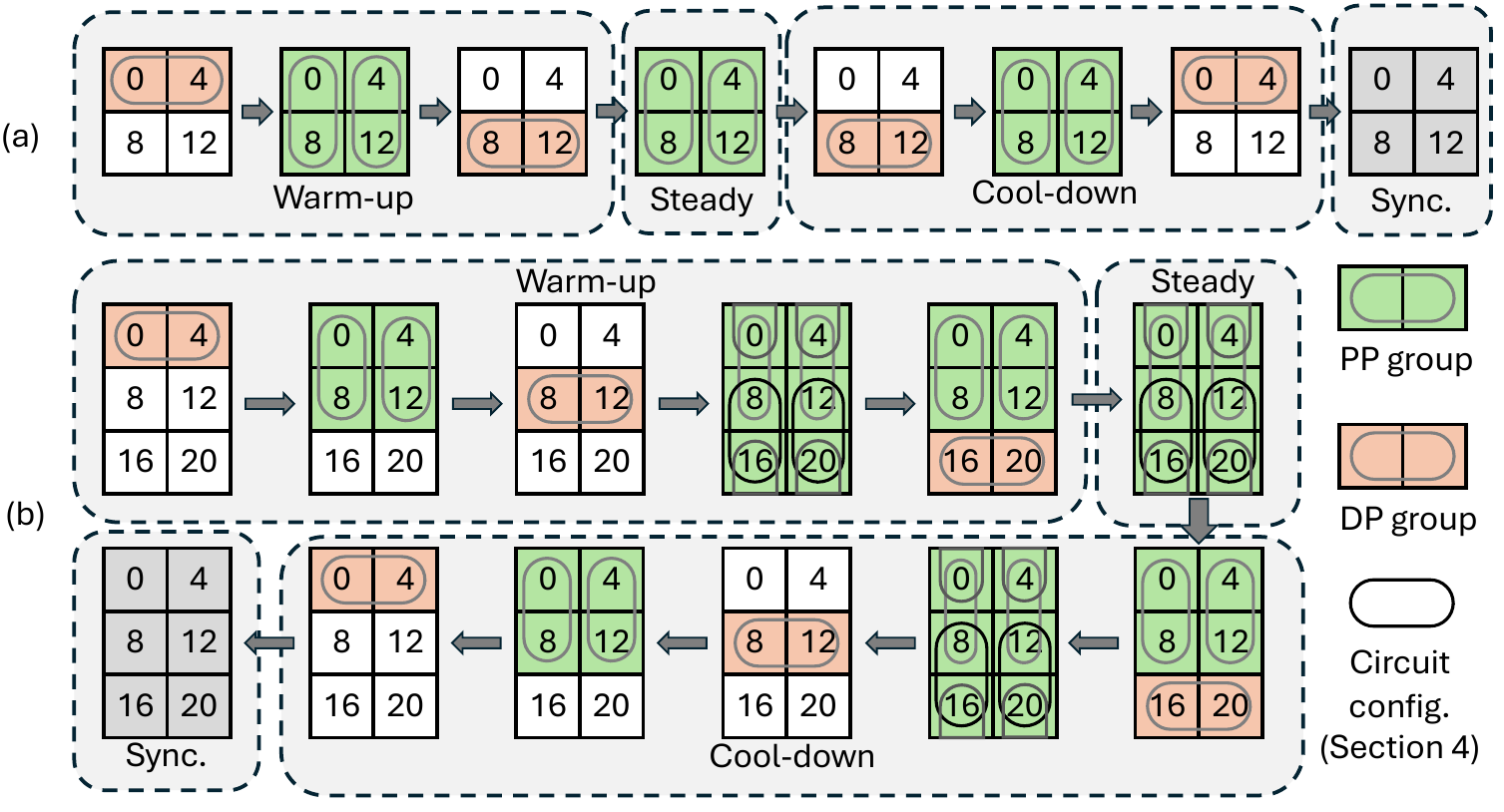}
    \caption{Communication pattern for PP and FSDP in one iteration, split based on the warm-up, steady, and cool-down stages of the pipeline (4 rails in total, only rail 0 is shown, TP is hidden). (a) PP=2, FSDP=2. (b) PP=3, FSDP=2. }
    \label{fig:config}
    \vspace{-0.5cm}
\end{figure}

\myparab{Understanding the communication pattern.} Figure~\ref{fig:config}(a) shows the communication pattern in a 3D-parallel workload. Stage 0 first performs micro-batch 0 forward pass (overlapped with per-layer \allgather to collect the next layer's parameters), and sends the activation to stage 1 hosted by rank 8 through a \sendrecv call along the pipeline dimension. Once the \sendrecv call is finished, rank 8 computes the forward pass, while doing \allgather. Then, it performs backpropagation for micro-batch 0, followed by pipeline \sendrecv. \reducescatter calls are issued after partial gradients are updated. During the optimizer step, several short \allreduce calls are issued for synchronization. We observe that the DP traffic does not overlap with PP traffic. Figure~\ref{fig:config}(b) shows the pattern for PP=3. The data dependency between operations, and PyTorch's lazy DTensor operation (\eg the first \allgather call for stage 1 only starts when it receives the activation from stage 0), dictate the sequential order between PP and DP traffic, though collectives from two dimensions are issued in different CUDA streams. We can then define the window as the idle time between two consecutive parallelism phases $P1$ and $P2$, which are two distinctive sets of communication groups: 
$$T_{window} = \min_{comm_j \in P2} T_{comm_j\_start}  - \max_{comm_i \in P1} T_{comm_i\_end},$$ where $comm_j \neq comm_i$ for all $comm_i \in P1$. In addition, 
$$T_{comm_j\_start} = \max_{rank_x \in comm_j} T_{rank_x\_comm_j\_start},$$
where $rank_x$ participates in $comm_j$, since the collective starts only when the slowest rank joins. $T_{comm_i\_end}$ is the end time of the $comm_i$, the same for all participating ranks.



\begin{figure*}[h!]
    \centering
    \begin{subfigure}{0.22\linewidth}
        \centering
        \includegraphics[width=\linewidth]{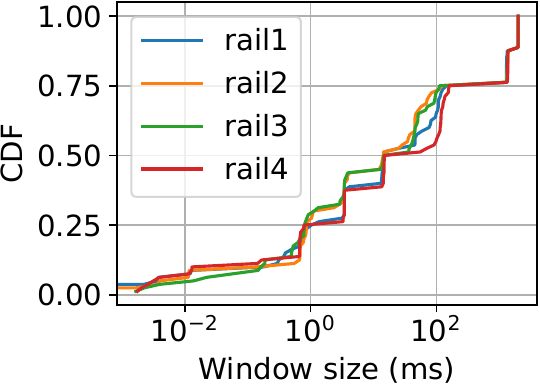}
        \caption{}
    \end{subfigure}
    \hfill
    \begin{subfigure}{0.22\linewidth}
        \centering
        \includegraphics[width=\linewidth]{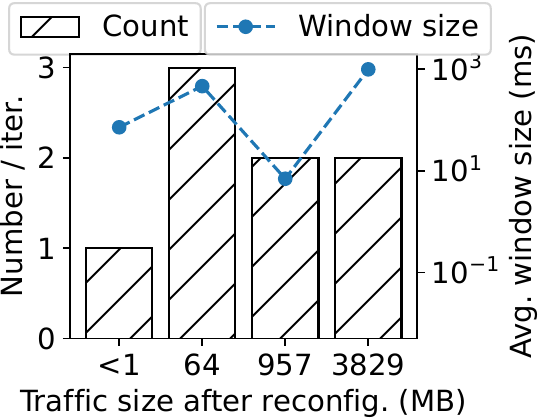}
        \caption{}
    \end{subfigure}
    \hfill
    \begin{subfigure}{0.19\linewidth}
        \centering
        \includegraphics[width=\linewidth]{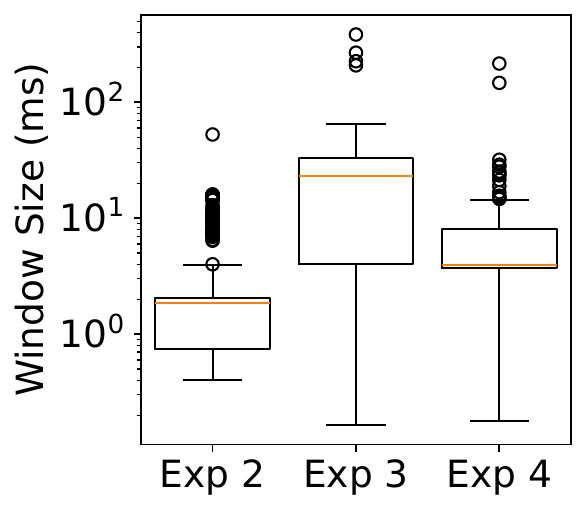}
        \caption{}
    \end{subfigure}
    \hfill
    \begin{subfigure}{0.25\linewidth}
        \centering
        \scriptsize
        \setlength{\tabcolsep}{2.2pt}
        \renewcommand{\arraystretch}{1.25}
        \resizebox{\linewidth}{!}{%
        \begin{tabular}{rrrrrr}
        \toprule
        \textbf{GPUs} & \textbf{DP} &
        \makecell{\textbf{Mean}\\\textbf{(ms)}} &
        \makecell{\textbf{$p_5$}\\\textbf{(ms)}} &
        \makecell{\textbf{$p_{50}$}\\\textbf{(ms)}} &
        \makecell{\textbf{$p_{95}$}\\\textbf{(ms)}} \\
        \midrule
        2,048 & 16  & 670  & 47.4 & 187  & 3,988 \\
        4,096 & 32  & 336  & 23.2 & 94.0 & 2,003 \\
        8,192 & 64  & 169  & 10.7 & 47.3 & 1,010 \\
        16,384 & 128 & 85.2 & 3.78 & 23.9 & 513 \\
        \bottomrule
        \end{tabular}}
        \caption{}
    \end{subfigure}

    \caption{(a) CDF of window size from 10 iterations in Exp. 1. (b) Rail 0 window breakdown based on traffic volume after the window and before the next window, in one iteration of Exp. 1. <1MB: \allreduce synchronization calls, 64MB: PP \sendrecv, 957MB: DP \allgather, 3829MB: DP \reducescatter. (c) Rail 0 window-size box plot for the step latency for Exp. 2--4\del{from three experiments}. (d) \rev{Window size statistics for a GPT-style workload~\cite{man2025stage} under DP strong scaling at frontier scale running on B200.}} 
    \label{fig:window}
\end{figure*}

\rev{
\begin{figure}[t]
    \centering
    \includegraphics[width=0.8\linewidth]{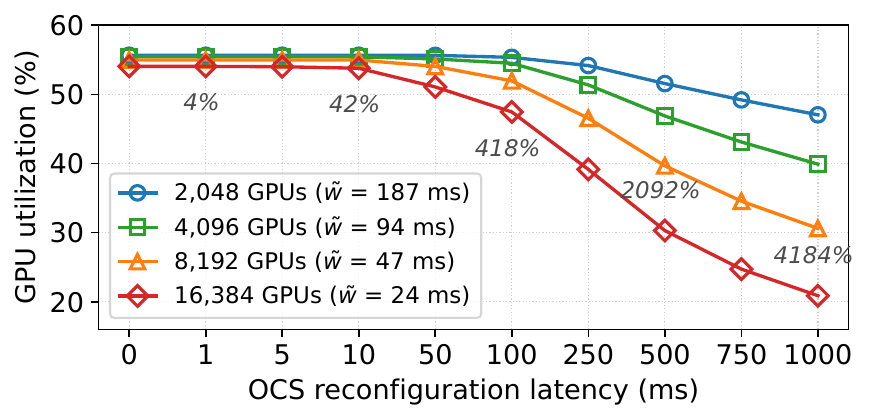}
    \caption{\rev{GPU utilization vs.\ OCS reconfiguration latency for a frontier workload on B200 (Figure~\ref{fig:window}(d)). 
    $\tilde{w}$ is a workload's median window. Percentages show the latency as a fraction of the median window at the largest (16{,}384-GPU) scale.}}
    \label{fig:sim_util_window}
\end{figure}
}

\myparab{Empirical findings.} To study the window sizes, we run an LLM training workload using TorchTitan~\cite{liang2024torchtitan} on the Perlmutter supercomputer~\cite{nersc_perlmutter_architecture}.
\del{The nodes are connected by Slingshot 11 interconnect fabric~\cite{hpe_cray_ex}. Each node has 4 A100 GPUs inter-connected via NVlink 3.0.}
We conduct \del{three}\rev{four} experiments: \textbf{(Exp. 1)} Llama-3-8B with TP=4 (intra-node), FSDP=2, PP=2, global batch size=16, sequence length=8192. \rev{\textbf{(Exp. 2)} DeepSeek-V3-236B with TP=4, FSDP=8, PP=8, EP=32, global batch size=64, sequence length=1024.} \textbf{(Exp. 3)} Llama-3-8B with TP=4, FSDP=8, PP=2, global batch size=64, sequence length=8192. \textbf{(Exp. 4)} Llama-3-70B with TP=4, FSDP=4, PP=8, global batch size=32, sequence length=1024. \del{The PP schedule is 1-forward-1-backward~\cite{shoeybi2019megatron}}

We find that reconfiguration windows are typically millisecond-scale. Figure~\ref{fig:window}(a,b) plots the CDF of window durations for the first workload and groups windows by the traffic volume of the following communication operation $comm_j$. More than 75\% of windows are longer than 1\,ms, and the distributions are similar across rails. The highest-volume operation, \reducescatter, is preceded by the longest windows, averaging roughly 1,000\,ms. For the workloads in Figure~\ref{fig:window}(c), \rev{the median window duration \rev{changes} 
because of varying collective sizes and different parallelism configurations. Even so, the median duration remains above 1\,ms.}
\del{the average window duraction decrease because longer communication operations delay the start of subsequent windows and denser model compute graphs introduce more communication operations. Even so, the average duration remains above 1\,ms.}

\rev{To understand the window duration in larger deployments, we simulate a frontier-scale workload under strong scaling, with a large global batch during frontier model pre-training (GPT-style~\cite{man2025stage}, PP=4, TP=32, global batch size 15,360, sequence length 4096~\cite{liu2024deepseek}, 96 layers, hidden dimension 8192) on B200 architecture using ASTRA-sim~\cite{astrasim} (Figure~\ref{fig:window}(d)). 
Compared to previous experiments, the larger local batches increase computation load, lengthening the windows.
The window size decreases under strong scaling:
the median window duration falls from 187\,ms at 2,048 GPUs to 23.9\,ms at 16,384 GPUs.
We further estimate and plot the GPU utilization (fraction of time spent on computation), assuming network reconfiguration is performed at the beginning of each window in Figure~\ref{fig:sim_util_window}:
\begin{equation}
\label{eq:gpu-util}
\begin{aligned}
\mathit{Util}\;(\%) &=
\frac{T_{\mathit{compute}}}
{T_{\mathit{compute}} + T_{\mathit{non\_overlap\_comm}} + T_{\mathit{stall}}}
\times 100, \\
T_{\mathit{stall}} &= \sum_i \max(0,\; T_{\mathit{reconfig}} - T_{\mathit{window},i}).
\end{aligned}
\end{equation}
Here, $T_{\mathit{non\_overlap\_comm}}$ is the non-overlapped communication time, $T_{\mathit{reconfig}}$ is the reconfiguration latency, and $T_{\mathit{window},i}$ is the duration for window $i$. Utilization is flat while the reconfiguration latency stays below most of the windows and degrades due to slower switches or larger deployments. With fast OCSes ($T_{\mathit{reconfig}}<10$\,ms)~\cite{sipac,neye}, the GPU utilization remains high for frontier model training. Additional results for smaller batch sizes (\eg 3,072 used in the initial pre-training phase~\cite{liu2024deepseek}) are included in Appendix~\S\ref{sec:window-batch}.} 
For general workloads, the number of windows in one iteration can be determined by Equation~\ref{eq:wndcnt}, assuming FSDP is used and the TP domain does not exceed scale-up. In the equation, $n_{\text{layer}}$ is the number of model layers, $n_{\text{microbatch}}$ is the number of microbatches per global batch, and $PP$ is the pipeline parallel degree.
\del{Using the training configurations reported by~\cite{chu2025scaling}, there are 127 windows over one Llama3.1-405B training iteration, approximately 20 seconds with 1k H100s ($\approx 6$ windows/second)~\cite{nvidia2025llama31_405b_dgxc}.}

\myparab{Summary.}
The non-overlapping communication pattern across parallelism dimensions creates an opportunity where each time window between parallelism phases can be used to configure circuits tailored to the upcoming collective operation. The size of individual windows depends on multiple factors, including the model size, batch size, sequence length, number of layers per pipeline stage, \rev{parallelism dimensions} and communication group size. Finally, our findings indicate that circuit configuration can have limited impact on ML application performance if the reconfiguration delay is on the order of milliseconds, allowing it to be hidden within these natural idle periods between parallelism phases.

\begin{equation}
\label{eq:wndcnt}
\resizebox{\columnwidth}{!}{$
\begin{aligned}
\text{Window count} \leq&
\underbrace{4 \cdot (PP - 1)}_{\substack{\text{PP and FSDP}\\\text{fwd/bwd interleave}}}
+ \underbrace{2 \cdot \frac{n_{\text{layer}}}{PP} - 1}_{\substack{\text{CP/EP and FSDP}\\\text{1st microbatch fwd interleave}}}
+ \underbrace{4 \cdot n_{\text{microbatch}}}_{\substack{\text{CP/EP and PP}\\\text{fwd/bwd interleave}}} \\
&+ \underbrace{2 \cdot n_{\text{microbatch}} \cdot \left(2 \cdot \frac{n_{\text{layer}}}{PP} - 1\right)}_{\substack{\text{CP and EP}\\\text{fwd/bwd interleave}}}
+ \underbrace{4}_{\substack{\text{PP warm-up, steady,}\\\text{cool-down, sync., and}\\\text{state transition}}}
\end{aligned}
$}
\end{equation}

\section{Parallelism-driven Rail Reconfiguration}

\label{sec:design}

Photonic rails limit the connection degree of GPUs, in turn limiting effective hybrid parallelism (\S\ref{sec:challenges}). Therefore, to realize photonic rails in ML datacenters where jobs employ hybrid parallelisms, it is essential to reconfigure the scale-out optical fabric \emph{within} each job to ensure each parallelism phase has sufficient degree of connectivity.

\begin{figure}[h!]
    \centering
    \includegraphics[width=1\linewidth]{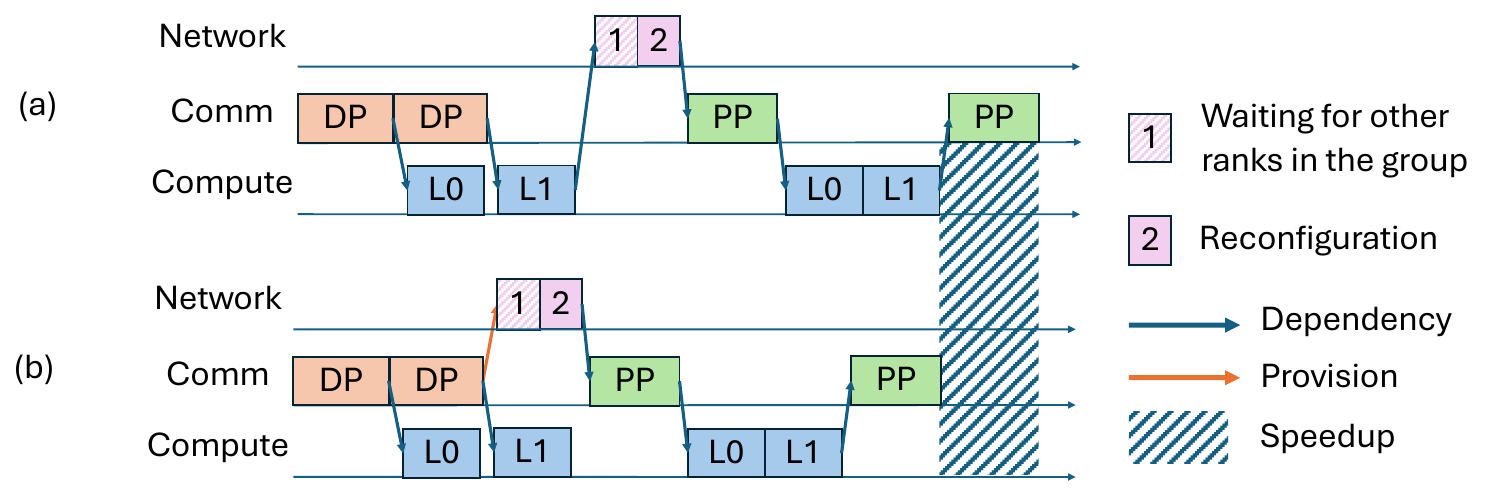}
    \caption{Reconfiguration during the warm-up stage of ranks 0 and 4, (a) without provisioning, (b) with provisioning.}
    \label{fig:demo}
\end{figure}

In-job topology reconfiguration must satisfy three objectives: (1) minimize reconfiguration delay, (2) reduce reconfiguration frequency to maximize circuit uptime, and (3) avoid conflicts between new circuits and ongoing traffic in the network. Meeting these objectives requires control logic that has knowledge of the parallelism structure of the ML workload. Therefore, this control logic must reside in the application layer, where parallelism-level hints are directly accessible. Moreover, by recognizing changes in parallelism phases (\eg DP $\rightarrow$ TP), the control logic can \emph{provision} circuits speculatively---initiating reconfiguration as soon as the previous kernel completes (Figure~\ref{fig:demo})---and suppress redundant reconfigurations when consecutive collectives belong to the same parallelism dimension (Figure~\ref{fig:config}). Note that our design choice to operate at the application layer departs from prior reconfigurable network controllers that operate on packets or flows at layers 2 and 3~\cite{cthrough,projector,farrington2010helios,amir2024shale,shoal,mellette2017rotornet,liang2024negotiator,ballani2020sirius}.

\begin{figure*}[h!]
       \centering
    \includegraphics[width=0.95\linewidth]{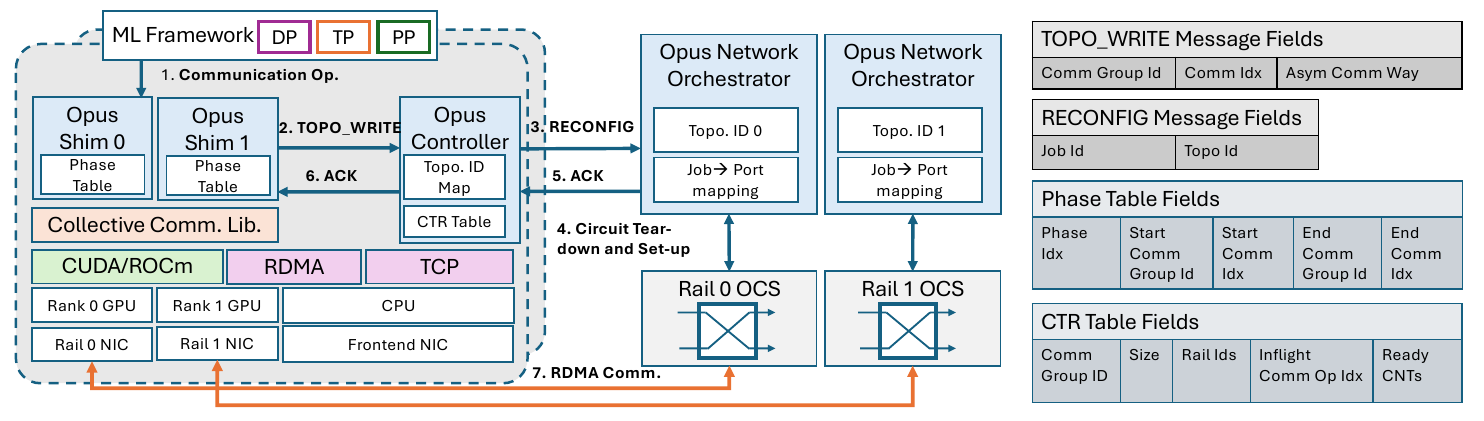}
       \caption{\sysname system architecture.}
       \label{fig:design}
\end{figure*}

\subsection{\sysname System Architecture}

We present \sysname, an application-layer control plane for parallelism-driven reconfiguration of optical circuit switches in rail-based datacenter fabrics. A distributed ML training job alternates between \emph{parallelism phases}: contiguous intervals during which all scale-out communication belongs to a single parallelism dimension (\eg all collectives are DP \allgather or \reducescatter, or all operations are PP \sendrecv). Phase boundaries, \ie transitions like PP $\rightarrow$ DP, are the natural points at which the rail's circuit topology can be safely reconfigured without disrupting in-flight traffic.

\sysname exploits this structure by treating rail connectivity as an allocatable resource whose shape is derived from the upcoming collective's communication pattern, and reconfigures the optical fabric \del{only at phase boundaries}\rev{accordingly}. To achieve this, \sysname introduces a control layer between ML training frameworks and collective communication libraries, comprising three components (Figure~\ref{fig:design}): \rev{the \textbf{\sysname shim}, which runs on every GPU rank within a job and intercepts collective calls, detects phase boundaries, and issues reconfiguration requests; the \textbf{\sysname controller}, spawned once per job, which synchronizes reconfiguration demands across ranks and forwards topology requests to the network; and the \textbf{\sysname network orchestrator}, spawned once per rail, which translates topology requests into OCS port-to-port programming commands.}

\del{
\begin{compactitem}
\item The \myparab{\sysname shim} runs on every GPU rank within a job.
  It intercepts collective calls, detects phase boundaries, and issues
  reconfiguration requests.
\item The \myparab{\sysname controller} is spawned once per job. It
  synchronizes reconfiguration demands across ranks and forwards
  topology requests to the network.
\item The \myparab{\sysname network orchestrator} is spawned once per
  rail. It translates topology requests into OCS port-to-port
  programming commands.
\end{compactitem}}

In \sysname, the data path for inter-GPU communication is simple: GPU $\rightarrow$ NIC $\rightarrow$optical fabric $\rightarrow$ NIC $\rightarrow$ GPU. \sysname's \emph{control path} decides which physical circuits are active at any instant. This separation keeps the datapath free of packet switching while allowing control logic to exploit parallelism-level hints available in the ML stack. Together, the shim and controller present an \emph{illusion of all-to-all} GPU connectivity. 

\myparab{The \sysname Shim.}
One shim instance runs per GPU rank, so the number of shim instances
equals the total number of GPUs in the job. Each shim maintains:
(i)~a \emph{phase table}, populated during profiling, that records the sequence of parallelism
phases and the communication operations that demarcate them;
(ii)~an operation index $\mathit{idx}$ that tracks progress through
the communication schedule; and (iii)~the current parallelism phase.
Each shim handles all $N_{\mathit{parallel}}$ communication groups to
which its GPU belongs.

The shim intercepts every collective communication call from the
distributed ML framework and extracts per-call metadata: the
\emph{communication group} (participant ranks and their ordering),
the \emph{operation type} (collectives like \allgather and
\allreduce, or point-to-point operations like \sendrecv), and
the \emph{traffic size}. Using this metadata, the shim classifies the operation as either a management operation (\eg a barrier at the end of a forward-backward
step) or a data-carrying operation, and decides: \rev{(1) \textbf{Network selection}—RDMA GPU backend
(scale-out) or CPU frontend network (Algorithm~\ref{alg:pre-coll},
lines~3,~16); (2) \textbf{Reconfiguration trigger}—whether the
operation marks a new parallelism phase requiring a topology change
(Algorithm~\ref{alg:pre-coll}, line~7); and (3)
\textbf{Timing}—reconfigure \emph{before}
(on-demand, Algorithm~\ref{alg:pre-coll}, line~8) or \emph{after}
the preceding operation (provisioning, Algorithm~\ref{alg:post-coll},
line~3).}

\del{
\begin{compactitem}
\item \myparab{Network selection:} Should this operation use the RDMA
  GPU backend (scale-out) or the CPU frontend network
  (Algorithm~\ref{alg:pre-coll}, line~3 \& line~18.)?
\item \myparab{Reconfiguration trigger:} Does the upcoming operation
  belong to a different parallelism phase, requiring a topology
  change? (Algorithm~\ref{alg:pre-coll}, line~9.)
\item \myparab{Reconfiguration timing:} Should the topology be
  reconfigured \emph{before} the operation (on-demand,
  Algorithm~\ref{alg:pre-coll}, line~10) or \emph{after} the
  preceding operation completes (provisioning,
  Algorithm~\ref{alg:post-coll}, line~6)?
\end{compactitem}
}

When reconfiguration is needed, the shim issues a blocking
\texttt{topo\_write} call to the controller, carrying three fields:
$$(\mathit{comm\_group\_id},\; \mathit{idx},\;
\mathit{asym\_comm\_way}),$$ 
with $\mathit{asym\_comm\_way}$ indicating the way (or stage) in an asymmetrical communication group (\eg PP).

\del{
\begin{algorithm}[h!]
    \footnotesize
    \caption{Pre-communication Control Logic}
    \begin{algorithmic}[1]

    \Procedure{pre\_comm}{$comm\_op$}
        \If{$comm\_op$ is scale\_up \textbf{or} management}
            \State{select scale-up or CPU front-end network}
            \State \Return
        \EndIf
        
        \State{wait till topology is free}
        \State $shift \gets \Call{phase\_change\_before}{}$ 
        \If{$mode = \textsc{default}$}
            \If{$shift$ \textbf{or} $comm\_op$ is asymm.}
                \State \Call{topo\_write}{$comm\_group.id,\; idx,\; asym\_way$}
            \EndIf
        \EndIf
    
        \If{$shift$}
            \State $comm\_stage \gets comm\_stage + 1$
            \State set topology busy
        \EndIf
    
        \State $idx \gets idx + 1$
        \State{select GPU backend network}
        \State register \textsc{post\_comm} callback 
    \EndProcedure
    
    \end{algorithmic}
\end{algorithm}
}

\rev{
\begin{algorithm}[h]
    \caption{Pre-communication Control Logic}
    \label{alg:pre-coll}
    \begin{algorithmic}[1]
    \setlength{\itemsep}{0pt}
    
    \Procedure{pre\_comm}{$comm\_op$}
        \If{$comm\_op$ is scale\_up \textbf{or} management}
            \State select scale-up or CPU front-end network
            \State \Return
        \EndIf
        
        \State wait till topology is free
        \State $shift \gets \Call{phase\_change\_before}{}$ 
        \If{$mode = \textsc{default}$ \textbf{and} ($shift$ \textbf{or} $comm\_op$ is asymm.)}
            \State \Call{topo\_write}{$comm\_group.id,\; idx,\; asym\_way$}
        \EndIf
    
        \If{$shift$}
            \State $comm\_stage \gets comm\_stage + 1$
            \State set topology busy
        \EndIf
    
        \State $idx \gets idx + 1$
        \State select GPU backend
        \State register \textsc{post\_comm}
    \EndProcedure
    
    \end{algorithmic}
    
\end{algorithm}
}

\myparab{The \sysname Controller.}
One controller instance is spawned per job. It maintains a
\emph{CTR~Table} (Figure~\ref{fig:design}) with metadata for every
communication group: sockets to each shim, the group size, the
rail~IDs used by the group, the index of the in-flight communication
operation, and a \emph{ready counter} that tracks how many ranks have
issued a \texttt{topo\_write} for the current operation. 
For a job with three parallelism dimensions of sizes $P_1$, $P_2$,
$P_3$, there are $P_1 P_2 + P_2 P_3 + P_3 P_1$ communication groups
in total. The controller also maintains a per-rail topology identifier,
$\mathit{topo\_id}_i$, which encodes the current connectivity
requirement of the job on rail~$i$. 

\myparab{Topology ID encoding.}
The $\mathit{topo\_id}$ is a decimal integer whose digit positions
correspond to the \emph{ways} (stages) of the job's asymmetrical
parallelism (\eg the $P_{\mathit{asym}}$ stages of PP). Each digit
value encodes which parallelism dimension currently ``owns'' the
connectivity for that stage:
$0$ denotes PP (asymmetrical), and digits $1, 2, \ldots$ denote
symmetrical parallelisms (DP, CP, EP, \etc). Up to~9 symmetrical
parallelisms can be encoded per digit.

\begin{figure}[t]
    \centering
    \includegraphics[width=1\linewidth]{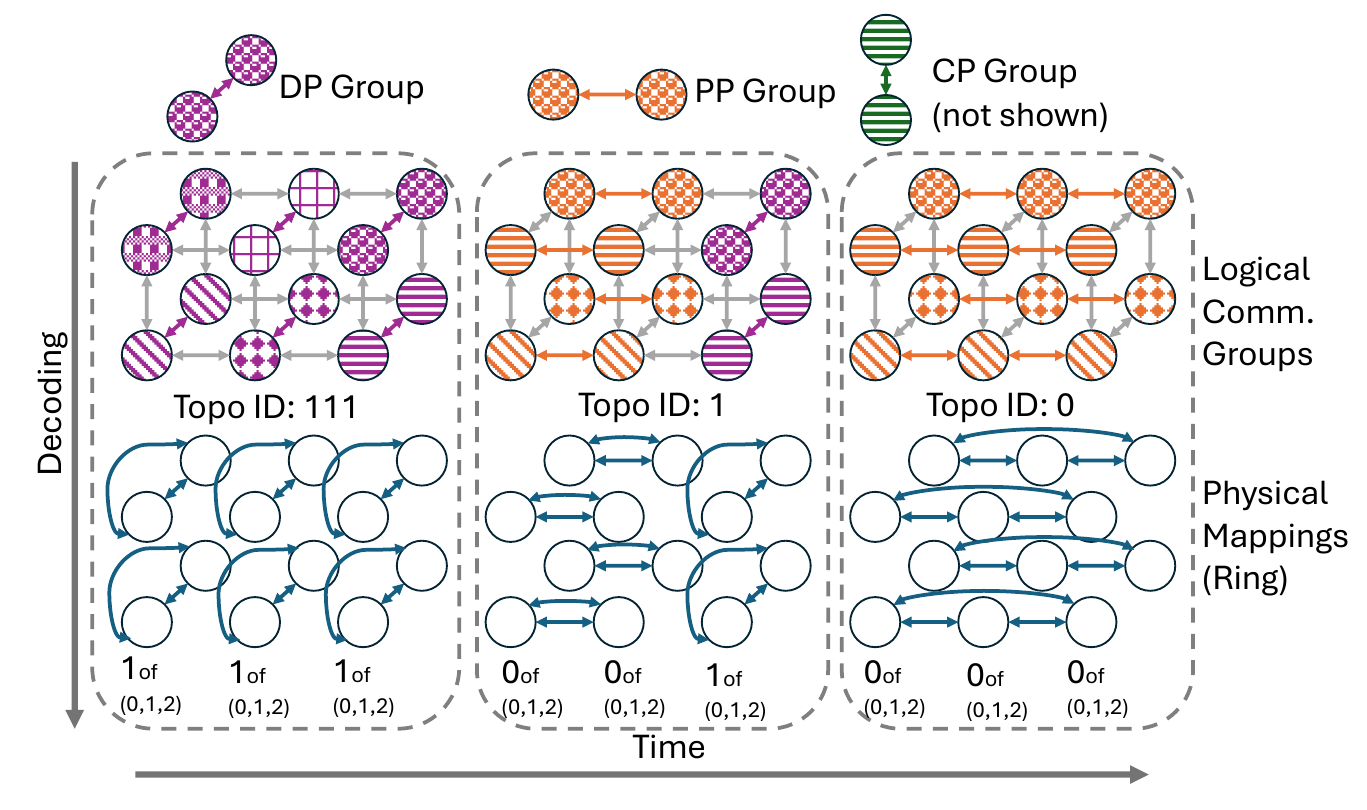}
    \caption{Translating parallelism shifts to topology reconfigurations for a workload with PP=3, DP=2, CP=2. PP is the asymmetrical parallelism, and DP \& CP are symmetrical parallelisms. \sysname topology orchestrator updates sub-mappings of affected ways in PP when the topology needs to be reconfigured. The mapping constructs per-group rings.}
    \label{fig:mapping}
    \vspace{-0.3cm}
\end{figure}

\myparab{Example (Figure~\ref{fig:mapping}).}
Consider a job with PP$=$3, DP$=$2, CP$=$2. The PP dimension is
asymmetrical, so the $\mathit{topo\_id}$ has three digits—one per
pipeline stage. When all three stages perform DP \allreduce, the
topology is $\mathit{topo\_id} = 111$ (all digits set to~1,
representing DP). When stage~0 and~1 transition to PP \sendrecv while
stage~2 remains on DP, the digits for stage~0 and~1 toggle to~0,
yielding $\mathit{topo\_id} = 001$. The orchestrator detects the
changed digits and reprograms only the sub-mapping for stage~0 and ~1.

\myparab{Runtime synchronization.}
The controller serves as a synchronization barrier. When a rank
joins with a \texttt{topo\_write}, the controller
increments the ready counter for the corresponding communication
group and operation index.
Once the counter reaches the group size, the controller:
(1)~updates the $\mathit{topo\_id}$ based on
$\mathit{asym\_comm\_way}$ and $\mathit{comm\_group\_id}$;
(2)~forwards the updated $\mathit{topo\_id}$ to one or more network
orchestrators;
(3)~waits for \texttt{ACK}s confirming reconfiguration is complete;
(4)~\texttt{ACK}s all ranks in the group; and
(5)~clears the ready counter for operation $\mathit{idx}$.

\myparab{The \sysname Network Orchestrator.}
One orchestrator instance manages each rail's OCSes. For every job $j$
on rail~$i$, the orchestrator stores the current $\mathit{topo\_id}$
and the port assignments for that job.

\myparab{Sub-mapping decomposition.}
A na\"ive approach stores all possible port-to-port mappings for every
job, requiring
$O\bigl(\sum_{j} N_{\mathit{parallel},j}^{P_{\mathit{asym},j}} \cdot
N_{\mathit{rank},j}\bigr)$
space, where $N_{\mathit{parallel},j}$ is the number of parallelism
dimensions, $P_{\mathit{asym},j}$ is the asymmetrical parallelism
degree (\eg PP stage count), and $N_{\mathit{rank},j}$ is the number
of ranks of job~$j$ on this rail.
\sysname reduces this by decomposing the mapping into
$P_{\mathit{asym},j}$ \emph{sub-mappings}—one per pipeline stage—each
covering $N_{\mathit{rank},j} / P_{\mathit{asym},j}$ ports. The
total storage becomes
$O\bigl(\sum_{j} N_{\mathit{parallel},j} \cdot
N_{\mathit{rank},j}\bigr)$,
and each reconfiguration event programs only
$O(N_{\mathit{rank},j} / P_{\mathit{asym},j})$ ports rather than
all $N_{\mathit{rank},j}$.

\myparab{Reconfiguration dispatch.}
Upon receiving a $\mathit{topo\_id}$ from the controller, the
orchestrator compares each digit against the current
$\mathit{topo\_id}$ for that job. A changed digit triggers
reprogramming of the corresponding sub-mapping (Figure~\ref{fig:mapping}).

Two cases can arise: (i)~\emph{symmetrical-to-asymmetrical shift} (\allgather in CP followed by \sendrecv in PP): at most two digits change, and the two corresponding
sub-mappings are rewired along the PP dimension. (ii)~\emph{asymmetrical-to-symmetrical shift} (\eg \sendrecv in PP followed by \allreduce in DP at stage~$m$): the orchestrator shifts sub-mapping~$m$ \emph{and} the sub-mappings of the stages that were
previously connected to~$m$. Sub-mappings belonging to other jobs, or sub-mappings of
the same job that are not involved in the current transition, remain
undisturbed. A non-blocking OCS, like the one we use in our hardware experiments, programs only the affected ports without impacting other circuits.

\subsection{Reconfiguration Protocol}
\label{sec:protocol}

\sysname provides two safety guarantees (\textbf{G}) and two performance
optimizations (\textbf{O}): 

\begin{compactitem}
\item \myparab{G1.} A communication operation is never carried on a
  topology that is under reconfiguration.
\item \myparab{G2.} A reconfiguration is never initiated while a
  communication operation is in flight on the affected sub-mappings.
\item \myparab{O1.} Number of reconfigurations per iteration is
  minimized.
\item \myparab{O2.} Reconfiguration delay is hidden whenever possible.
\end{compactitem}

\myparab{Event ordering.}
\del{Guarantees}G1 and G2 prevent packet loss. \del{\sysname enforces these guarantees}\rev{They are enforced} through a lock in the network backend. When the shim detects a phase boundary after a communication operation, it acquires the lock, issues a
\texttt{topo\_write}, and releases the lock only after receiving
\texttt{ACK} from the controller. No communication operation that uses GPU backend network can
proceed on the locked shim until the lock is released.

\myparab{Profiling parallelism phases.}
Optimization O1 requires knowing the parallelism phase structure in
advance. During the first several training iterations (\eg 5 steps),
the shim profiles the complete traffic pattern by recording the
sequence of collectives, their parallelism associations, and
inter-collective timing. From this trace, the shim identifies phase
boundaries—transitions between communication groups belonging to
different parallelism dimensions—and populates the phase table.
After profiling, the shim issues reconfiguration requests only at
phase boundaries, suppressing redundant reconfigurations when
consecutive operations belong to the same phase. The phase table
also enables locking (the shim knows whether the next operation
requires a different topology) and provisioning, which we describe next.

\rev{
\begin{algorithm}[t]
    \caption{Post-communication Control Logic}
    \label{alg:post-coll}
    \begin{algorithmic}[1]
    \setlength{\itemsep}{0pt}
    
    \Procedure{post\_comm}{$comm\_group,\; comm\_op$}
        \State $shift \gets \Call{phase\_change\_after}{}$ 
        \If{$mode = \textsc{provisioning}$ \textbf{and} ($shift$ \textbf{or} $comm\_op$ is asymm.)}
            \State $(n\_group\_id,\; n\_idx) \gets \Call{get\_next\_comm}{}$
            \State \Call{topo\_write}{$n\_group\_id,\; n\_idx,\; asym\_way$}
        \EndIf
    
        \If{$shift$}
            \State set topology free
        \EndIf
    \EndProcedure
    
    \end{algorithmic}
\end{algorithm}
}

\myparab{Provisioning.}
Optimization O2 exploits the predictability of the communication
schedule. In provisioning mode, the shim issues speculative
reconfiguration requests immediately after the \emph{last} operation
of the current phase completes (Algorithm~\ref{alg:post-coll},
line~5), rather than waiting until the \emph{first} operation of the
next phase arrives (Figure~\ref{fig:demo}). This allows the
orchestrator to program the OCS during the idle window between
phases.
The total delay exposed to communication across an iteration is
$\sum_i \max(0,\; T_{\mathit{reconfig}} - T_{\mathit{window},i})$.\del{,
where $T_{\mathit{reconfig}}$ is the reconfiguration duration and
$T_{\mathit{window},i}$ is the duration of idle window $i$.} Windows
with $T_{\mathit{window},i} \ge T_{\mathit{reconfig}}$ fully hide
reconfiguration and contribute no delay.
As shown in \S\ref{sec:challenges}, \del{over 75\%} \rev{most} phase-transition
windows exceed 1\,ms, \del{comfortably}accommodating OCS reconfiguration
times of \del{many}\rev{fast} switching technologies~\cite{sipac, neye}.

\myparab{Handling asymmetrical parallelism.}
For symmetrical parallelisms (\eg DP, CP, EP), all ranks in a
communication group participate in every collective, so locking and
provisioning operate at phase granularity: the shim locks before the
first operation of a new phase and unlocks after the last.
PP is \emph{asymmetrical}: different pipeline
stages may execute different parallelism phases concurrently. For
example, stage~0 may be processing a forward pass with DP
\reducescatter while stage~1 is idle, waiting for activations.

This heterogeneity means that a single per-phase lock is insufficient.
Instead, for PP communication, the \sysname shim performs locking and
provisioning at \emph{per-operation} granularity: every PP \sendrecv
triggers a \texttt{topo\_write} from both the sending and receiving
stage, regardless of the previous communication pattern. This ensures
the reconfiguration intent is always conveyed to the OCS fabric, even
when one stage has been idle.
A reconfiguration triggered by PP \sendrecv changes at most two
digits in the $\mathit{topo\_id}$, corresponding to the two
communicating stages.

\del{
\begin{algorithm}[h!]
    \caption{Post-communication Control Logic}
    \footnotesize
    \begin{algorithmic}[1]
    
    \Procedure{post\_comm}{$comm\_group,\; comm\_op$}
        \State $shift \gets \Call{phase\_change\_after}{}$ 
        \If{$mode = \textsc{provisioning}$}
            \If{$shift$ \textbf{or} $comm\_op$ is asymm.}
                \State $(n\_group\_id,\; n\_idx) \gets \Call{get\_next\_comm}{}$
                \State \Call{topo\_write}{$n\_group\_id,\; n\_idx,\; asym\_way$}
            \EndIf
        \EndIf
    
        \If{$shift$}
            \State set topology free
        \EndIf
    \EndProcedure
    
    \end{algorithmic}
\end{algorithm}
}

\rev{\myparab{Handling \alltoall.}
To handle the large fan-out in expert-parallel \alltoall, \sysname 
sets different topologies (ring orderings) across different rails. The topologies are 
chosen to maximize the connectivity and reduce forwarding cost in the 
scale-out network.
Concretely, \sysname uses divergent per-rail topologies so that their union
forms a maximally connected scale-out topology for each EP group, letting most pairs
exchange in a single scale-out hop through forwarding in the high-bandwidth scale-up domain.
Details about the topology design and the
routing protocol are in Appendix~\S\ref{sec:alltoall-topo}.
}

\myparab{Handling communication faults.}
\sysname handles transient failures through timeout-based detection.
If a reconfiguration \texttt{ACK} is not received within a configurable timeout, the controller retries. 
Persistent failures trigger fallback to a \emph{giant ring}—a static
circuit connecting all ranks—that provides basic connectivity at
reduced bandwidth. The training
framework is notified by the shim, and can checkpoint and restart affected ranks.
For OCS hardware failures, \sysname leverages rail redundancy: each
GPU connects to multiple rails through the scale-up interconnect, so
traffic can be rerouted through alternate rails at the cost of
increased contention.

We provide complete pseudocode in Appendix~\S\ref{sec:code}.

\subsection{Implementation}
\label{sec:impl}

We implement \sysname as a custom PyTorch distributed backend. A training job enables \sysname by passing \texttt{backend="opus"} to PyTorch's \texttt{init\_process\_group}, or equivalently, by setting a single configuration flag in frameworks built on PyTorch's
distributed API (\eg \texttt{enable\_opus\_backend} in TorchTitan~\cite{liang2024torchtitan}). No changes to model code or parallelism constructs (DP/PP/TP/CP/EP) are required to use \sysname. \sysname intercepts collective calls through PyTorch's \texttt{ProcessGroup} abstraction and routes them through the \sysname shim, which delegates data transport to NCCL.

\myparab{Shim runtime.}
The shim uses two threads to separate control-plane work from
data-plane execution. The main thread runs \texttt{pre\_comm} logic—network
selection, phase-shift detection, and reconfiguration
requests—before dispatching each collective to NCCL.
A dedicated callback thread runs \texttt{post\_comm} as soon
as the CUDA stream signals completion of a collective,
enabling provisioning to overlap with the idle
window between phases. In the common case (no phase shift), both threads execute
without blocking the data plane. When a phase transition occurs, the threads acquire a lock that serializes reconfiguration with communication (\S\ref{sec:design}).

\myparab{Traffic profiling.}
During the first five training steps, the shim profiles
the communication pattern by recording each collective's communication group and the operation index within a parallelism phase. The resulting traces are stored in the
\emph{phase table}, a cache that the shim consults in
subsequent steps to detect phase boundaries, suppress
redundant reconfigurations, and drive provisioning decisions.

\myparab{Network orchestrator interface.}
The orchestrator communicates with OCSes through a
vendor-neutral switch-driver interface supporting protocols
such as TL1, SCPI, and NETCONF. 
Configuration commands specify source--destination port
mappings. Hardware dependence is isolated behind this
interface, so the same controller logic can target a
production OCS or a software emulator during evaluation.

\myparab{Code.}
The shim runtime and controller comprise approximately
4{,}000 lines of C++, built against PyTorch~2.10.0, NCCL~2.28,
and CUDA~12.8. The network orchestrator for our hardware
testbed is 300 lines of Python, using a TL1 wrapper to
interface with the switch. We test \sysname with
TorchTitan~\cite{liang2024torchtitan} and provide simulation
tooling, built on top of ASTRA-sim~\cite{astrasim} for reconfigurable rail topologies. Both are open-sourced at \url{https://github.com/opusfabric/Opus}.
\section{Evaluating \sysname}
We evaluate \sysname across three complementary scales:
a small-scale hardware testbed
validates \sysname's reconfiguration mechanisms on real OCS
hardware; medium-scale emulation on the Perlmutter supercomputer
(up to 64 GPUs) measures end-to-end training overhead under
logically emulated circuit switches; and large-scale simulation in
Astra-Sim (up to 2{,}048 GPUs) with a custom reconfigurable network
backend evaluates \sysname's scalability and sensitivity to OCS
reconfiguration latency and link bandwidth.

\subsection{Lab Hardware Evaluation}
\label{sec:hw_eval}
\begin{figure*}[t]
    \centering
    \begin{subfigure}[b]{0.20\textwidth}
        \centering
        \includegraphics[width=\linewidth]{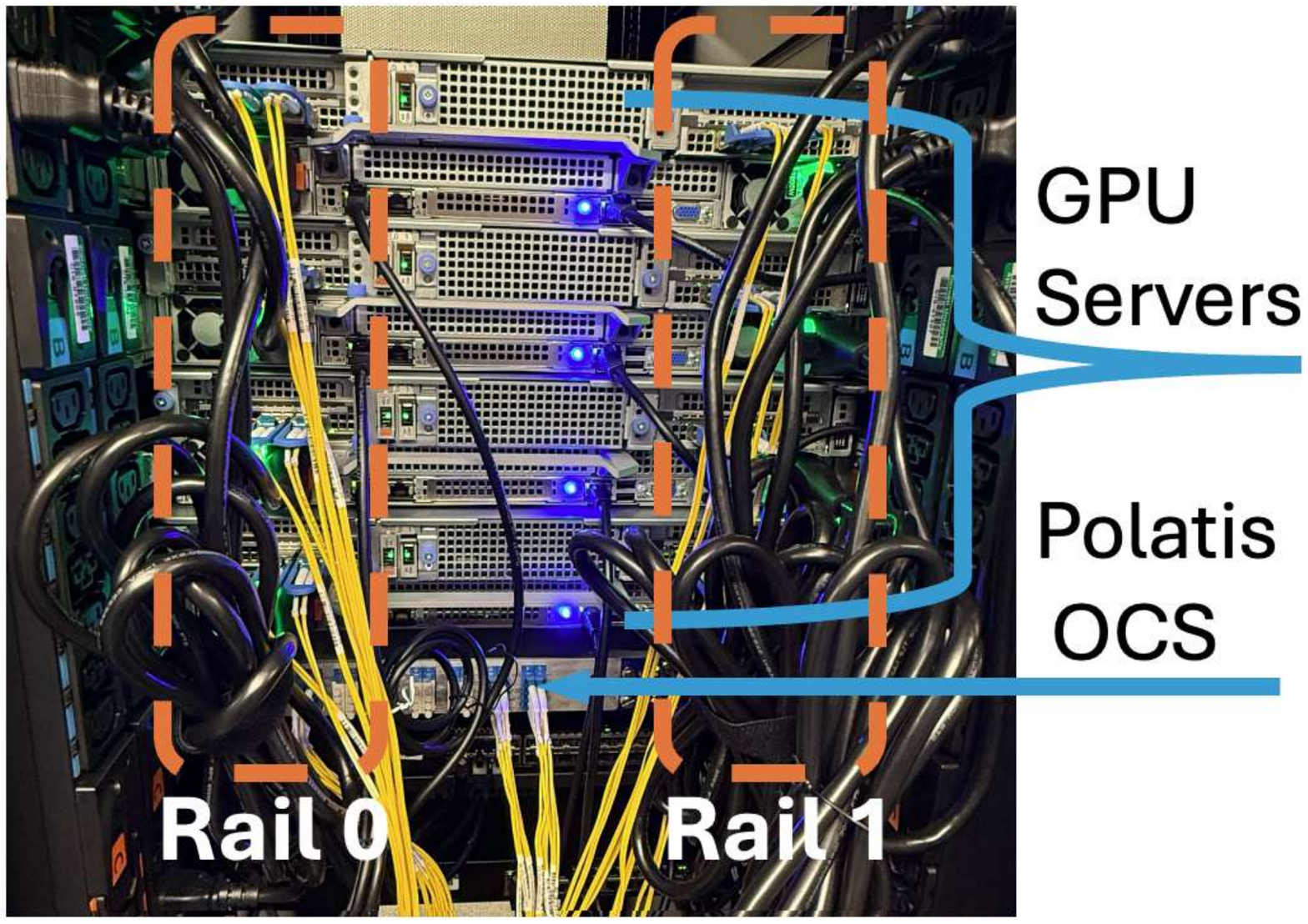}
        \caption{Testbed setup}
        \label{fig:testbed}
    \end{subfigure}
    \hfill
    \begin{subfigure}[b]{0.26\textwidth}
        \centering
        \includegraphics[width=\linewidth]{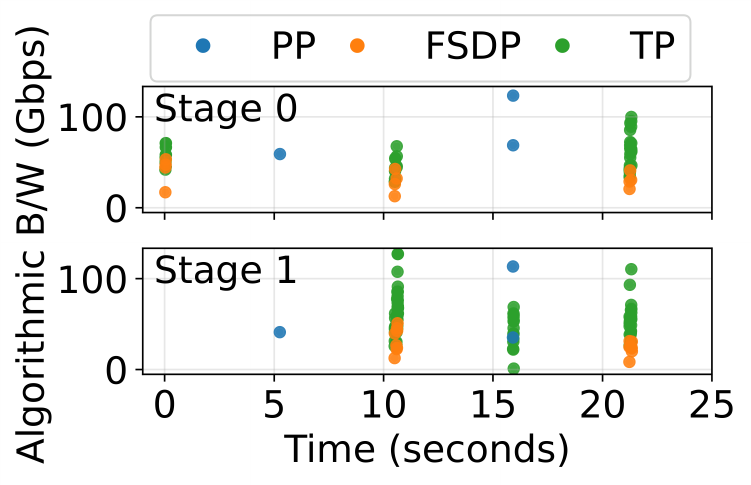}
        \caption{Training step communication}
        \label{fig:testbed_training}
    \end{subfigure}
    \hfill
    \begin{subfigure}[b]{0.26\textwidth}
        \centering
        \includegraphics[width=\linewidth]{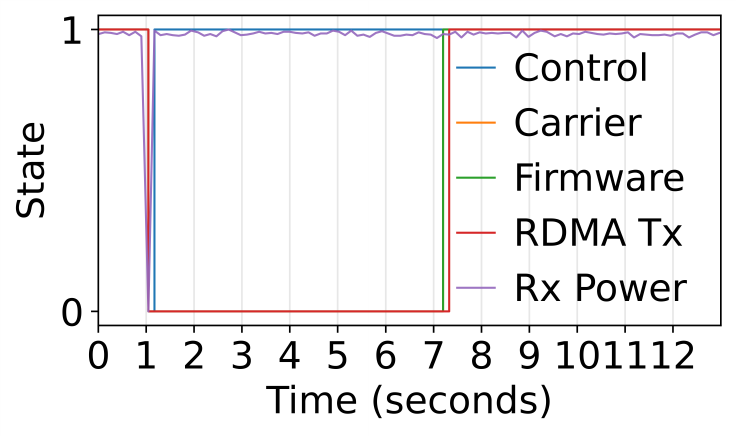}
        \caption{Reconfiguration timeline}
        \label{fig:timing_overall}
    \end{subfigure}
    \hfill
    \begin{subfigure}[b]{0.26\textwidth}
        \centering
        \includegraphics[width=\linewidth]{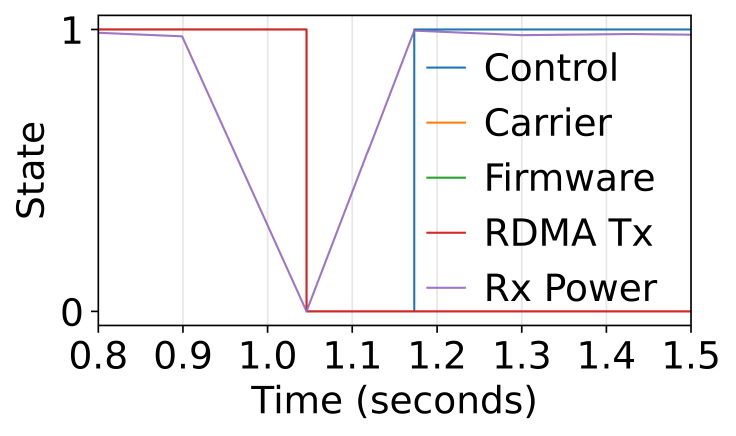}
        \caption{Zoomed timeline}
        \label{fig:timing_zoomed}
    \end{subfigure}
    \caption{Hardware testbed evaluation. (a) Physical testbed: 4 L40 GPU servers connected via a Polatis Series 6000 OCS forming two photonic rails. (b) Communication within one training step (Llama-3, DP=2, TP=2, PP=2) across two pipeline stages, showing \sysname\ reconfiguring between DP and PP phases. Algorithmic bandwidth is the collective buffer size divided by execution time. (c,d) Timeline of system signals during optical switch reconfiguration: Control is the control-plane command duration; Carrier is the Linux netdev carrier state; Firmware is the link status from \texttt{mlx5\_query\_vport\_state}; RDMA Tx indicates successful (1) or failed (0) sends; Rx Power shows normalized received optical power. The zoomed view (d) shows that the OCS and optical link recover within $\approx$200\,ms \rev{(with measurement overhead)}, but NIC firmware delays link-up by $\approx$6\,s.}
    \label{fig:hw_eval}
\end{figure*}

We evaluate \sysname on our hardware testbed (Figure~\ref{fig:hw_eval}(a)) that comprises one Polatis Series 6000 optical circuit switch with 64 optical LC/UPC ports~\cite{OCS_6000N} and four NVIDIA L40 GPU servers, each housing 2 L40 GPUs and 2 ConnectX-6 Dx Ethernet NICs. Each NIC exposes two 100G ports fitted with Cisco QSFP-100G-LR4 single-mode transceivers (1310\,nm), yielding 16 transceivers and 32 fiber cables connecting the servers to the switch in a two-rail topology. All components are commercially available and use standard networking interfaces and protocols. We use RoCEv2 for RDMA transport and NCCL 2.28 for collective communication. We made no modifications to the RoCEv2 protocol, Mellanox driver, or NCCL library. We set the RDMA \texttt{RETRY\_CNT} to 7 so that queue pairs do not enter an error state during reconfiguration events, allowing NCCL to reuse existing connections after circuits are reconfigured.

\myparab{Workload execution.}
We train a Llama-3 model (6 layers) with DP=2, TP=2, PP=2 on the hardware testbed with \sysname. Figure~\ref{fig:hw_eval}(b) shows the communication operations across two pipeline stages within one forward-backward step. TP collectives are handled by the intra-server PCIe interconnect and do not traverse the optical fabric. Note that on a production-grade testbed, the intra-server interconnect (\eg NVLink instead of PCIe) is much faster than the optical fabric. On the scale-out network, \sysname\ triggers 4 reconfiguration events per step, shifting the rail topology between DP and PP configurations. At each reconfiguration, \sysname\ allocates the full 200\,Gbps per-GPU bandwidth (2$\times$100G ports) to the active parallelism dimension, confirming that photonic rails can provide full-bandwidth, per-phase connectivity. 

\myparab{Reconfiguration timeline.}
We measure reconfiguration latency at each layer of the network stack to identify bottlenecks (Figure~\ref{fig:hw_eval}(c) and \ref{fig:hw_eval}(d)). \del{The \sysname control plane toggles the optical switch between two configurations in $\approx$200\,ms. Optical Rx power returns to baseline within $\approx$100 ms of the switch command}\rev{Optical Rx power returns to baseline within 200 ms of the switching command}, though \del{our 100 ms sampling interval}\rev{\texttt{ethtool}'s PHY-layer sampling interval (100 ms)} prevents resolving the exact \del{recovery}\rev{OCS switching} time\rev{~\cite{ethtool}}. Once the firmware reports link-up, the Linux kernel netdev carrier transitions within $\approx$70--80\,$\mu$s, and RDMA applications resume transmitting within an additional $\approx$100--200\,ms. \del{The transport and link layers therefore incur negligible overhead during reconfiguration, and no modifications to NCCL or the RoCEv2 stack are required.}

\myparab{Key Takeaway from hardware experiments.}
\begin{sloppypar}
The dominant delay in reconfiguring on our hardware testbed is in the NIC \rev{and transceiver} firmware: after Rx power stabilizes, the proprietary \del{Mellanox}firmware takes $\approx$6 seconds to report link-up via \texttt{mlx5\_query\_vport\_state}~\cite{nvidia_connectx6dx_firmware, nvidia_mlnx_en}. Disabling auto-negotiation (\texttt{PHY\_AUTO\_NEG} via \texttt{mlxconfig}~\cite{nvidia-mft} and \texttt{autoneg} via \texttt{ethtool}~\cite{ethtool}) reduces this to $\approx$3 seconds\del{, though rapid successive reconfigurations can increase variance up to $\approx$10 seconds. So we keep auto-negotiation permanently disabled}. This firmware bottleneck is not fundamental to photonic rails or \sysname---it reflects the firmware's assumption that link-state changes are rare failure events rather than routine reconfigurations. With firmware support for fast link-up~\cite{mellette2017rotornet}, the end-to-end reconfiguration latency would be bounded by the OCS switching time ($\approx$\del{200}\rev{25} ms for our Polatis switch\rev{~\cite{OCS_6000N}}, and $<$10 ms for state-of-the-art OCSes~\cite{neye}). \del{Our emulation and simulation results (\S\ref{sec:emulation}, \S\ref{sec:sim}) confirm that such latencies are well within the tolerance of large-scale ML training.}
\end{sloppypar}
\subsection{Emulation of \sysname on a Supercomputer}
\label{sec:emulation}

\begin{table}[t]
    \centering
    \small
    \begin{tabular}{ccccc}
    \hline
    \multicolumn{1}{p{0.5cm}}{\centering\textbf{Config}} & \textbf{Model} & \multicolumn{1}{p{1.6cm}}{\centering \textbf{Global\\Batch Size}} & \multicolumn{1}{p{0.8cm}}{\centering \textbf{Seq.\\ Length}} & \multicolumn{1}{p{2cm}}{\centering \textbf{Parallelism \\ (TP, FSDP, PP)}} \\
    \hline
    1 & Llama-3-8B & 16  & 8192  & (4, 2, 2) \\
    2 & Llama-3-8B      & 64  & 8192  &  (4, 8, 2)     \\
    3 & DeepSeek-V3-16B & 8   & 2048  &  (4, 1, 4)             \\ \hline
    \end{tabular}
    \caption{Workload configurations. \del{DP uses FSDP.}The number of microbatches equals the PP degree. Config.~3 uses only PP in the scale-out and requires no in-job reconfiguration.}
    \label{tab:llm-config}
\end{table}

To evaluate \sysname at larger scale without the NIC firmware bottleneck found in \S\ref{sec:hw_eval}, we emulate optical circuit switching on the Perlmutter supercomputer~\cite{nersc_perlmutter_architecture}. The \sysname shim and controller run unmodified. We replace the network orchestrators with logical circuit switches that (1) enforce one-to-one connectivity, (2) inject configurable reconfiguration delays, and (3) support non-blocking reconfiguration of a subset of circuits without disrupting the rest. Each scale-up domain has 4 NVIDIA A100 GPUs (4 rails) with 3rd-gen NVLink (800 Gbps per direction) and HPE Cray Cassini NICs (200 Gbps).

Our testing framework allows toggling \sysname on and off. With \sysname disabled, workloads run on the native PyTorch distributed backend with NCCL and GPUDirect RDMA over Perlmutter's Slingshot-11 fabric, which is a fully connected, electrical-packet-switched (EPS) network. This provides a direct comparison between \sysname on emulated photonic rails and the native software stack on ideal EPS. We evaluate three modes: (1) \emph{Native}: TorchTitan on EPS without \sysname, (2) \emph{\sysname}: emulated reconfigurable topology with configurable OCS delays, and (3) \emph{\sysname+Provisioning}: speculative reconfiguration of topology is enabled. Table~\ref{tab:llm-config} lists the workload configurations. All metrics are averaged over 5 training steps after 5 warm-up steps. The four topology orchestrators and the controller run on node~0. \rev{The emulation results do not reflect the congestion-free benefit of all-optical links.}

\begin{figure}[h!]
    \centering
    \includegraphics[width=\linewidth]{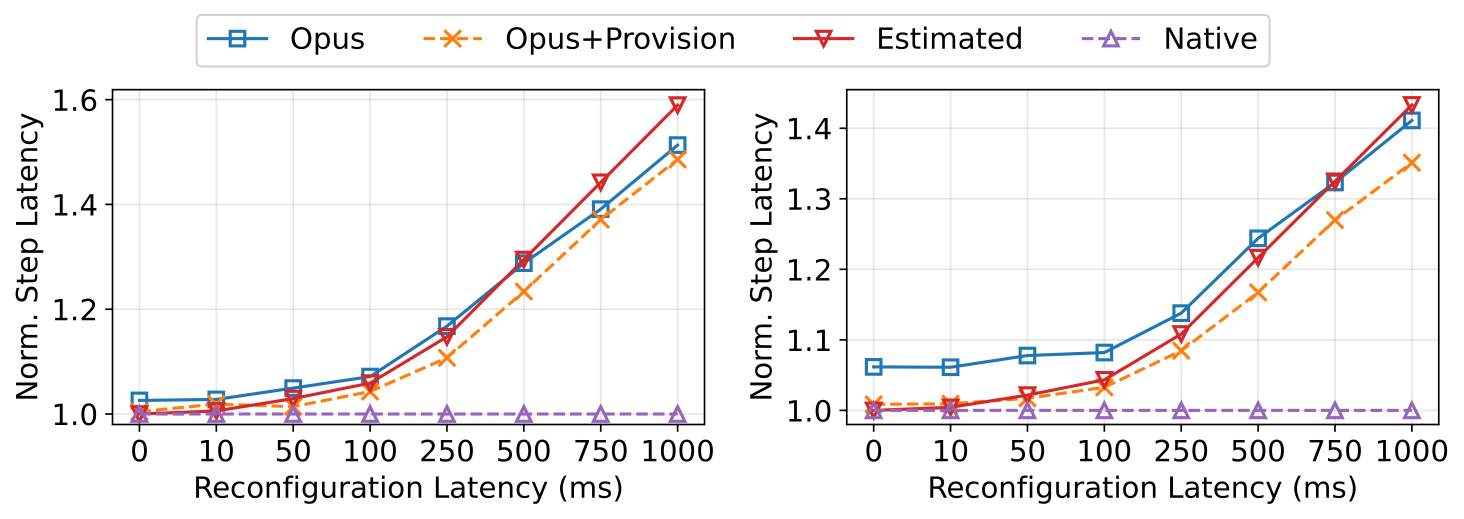}
    \caption{Step latency vs.\ emulated OCS reconfiguration latency for Config.~1 (left) and Config.~2 (right).}
    \label{fig:emulate_config}
\end{figure}

\myparab{Sensitivity to OCS reconfiguration latency.}
We sweep the emulated \del{OCS reconfiguration}latency from 0 to 1000 ms for Configs.~1 and~2 (Figure~\ref{fig:emulate_config}). Both configurations require 6 reconfigurations per training step. We compare against a simple analytical estimate: $T_\text{native} + T_\text{reconfig} \times N_\text{reconfig}$. At 50\,ms OCS latency, \sysname incurs 1.05$\times$ and 1.08$\times$ the native step time for Configs.~1 and~2 respectively. With provisioning, these overheads drop to 1.01$\times$ and 1.02$\times$, as speculative reconfiguration hides the OCS delay within the idle window between parallelism phases (\S\ref{sec:challenges}). The benefit of provisioning is more pronounced in Config.~2, where larger DP groups create wider inter-phase windows due to higher collective synchronization delays.


\begin{figure}[h]
\centering
    \includegraphics[width=\linewidth]{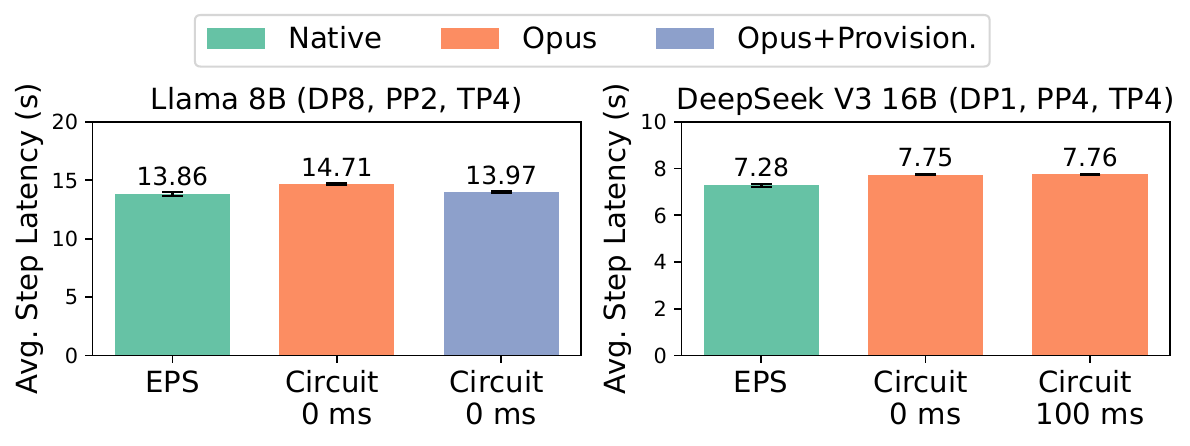}
    \caption{Control-plane overhead. (left) Config.~2 (64 GPUs): \sysname vs.\ \sysname+Provisioning vs.\ native EPS at 0\,ms emulated OCS latency. (right) Config.~3: \sysname at 0\,ms and 100\,ms OCS latency vs.\ native EPS. No reconfigurations occur, confirming \sysname suppresses unnecessary topology changes.}
    \label{fig:overhead_emulate}
\end{figure}

\myparab{Control-plane overhead.}
To isolate \sysname's control-plane overhead from OCS reconfiguration delay, we set the emulated latency to 0\,ms and measure Config.~2 at 64 GPUs (Figure~\ref{fig:overhead_emulate} (left)). The overhead---comprising \texttt{pre\_comm}/\texttt{post\_comm} logic, per-rail locking during phase transitions, and rank synchronization through the controller---is 6.13\%. With provisioning, the overhead drops to 0.79\%, because the synchronization cost is masked within inter-phase idle windows.

Figure~\ref{fig:overhead_emulate} (right) shows Config.~3, which places 4-way PP in the scale-out with no DP. Since all scale-out communication belongs to a single parallelism dimension, \sysname's topology orchestrator correctly suppresses all reconfigurations: the step latency is unchanged between 0\,ms and 100\,ms emulated OCS latency. The 6.46\% overhead is entirely due to \sysname's per-operation control logic for PP, which issues controller requests for every asymmetric Send/Recv (\S\ref{sec:protocol}). Also, the step duration does not increase with increased reconfiguration latency, showing that the \sysname topology orchestrator avoids unnecessary network reconfigurations.

\subsection{Large-scale Simulation of \sysname}
\label{sec:sim}

We evaluate \sysname at large scale using ASTRA-sim~\cite{astrasim}, a distributed ML system simulator, with Chakra ET~\cite{chakra} traces that replay realistic training iterations under hybrid parallelism. We simulate two \del{$\sim$80B}\rev{Llama/GPT-style} dense models (\del{Table~\ref{tab:sim_model}}\rev{hidden dimension 8192, feed-forward dimension 28672, 64 attention heads, 8 KV heads, 96 layers, sequence
length 4096, and global batch size 256})\del{at 64--2,048 GPUs}\rev{ and a DeepSeek-V3-style mixture-of-experts model,} and report end-to-end iteration time.

\myparab{Reconfigurable network backend.}
We extend ASTRA-sim with a \emph{reconfigurable analytical} network backend that supports time-varying connectivity. The backend consumes a set of candidate circuit configurations, each expressed as a directed bandwidth matrix indexed by a topology ID. Zeroes in the matrix represent absent circuits. The active matrix changes dynamically as \sysname selects configurations at runtime, while base link latency and reconfiguration latency are applied uniformly from a YAML configuration. To ensure correctness, the backend rejects reconfiguration requests while collectives are in flight or another reconfiguration is pending. Accepted reconfigurations drain active links before applying new settings. During the reconfiguration interval, arriving traffic queues on links and is released upon completion.

\del{
\begin{table}[h]
    \centering
    \footnotesize
    \begin{tabular}{ccccc}
        \hline
        \textbf{Model Type} & \textbf{dvocal} & \textbf{dmodel} & \textbf{dff} & \textbf{seq}\\
        \hline
        LLaMA/GPT & 32000 & 8192 & 28672 & 4096 \\
        \hline
        \textbf{head} & \textbf{kvhead} & \textbf{num\_stacks}& \textbf{batch} & \\
        \hline
        64 & 8 & 96 & 256 &  \\
        \hline
    \end{tabular}
    \caption{Specification for the simulated 80B GPT and LLaMA.}
    \label{tab:sim_model}
\end{table}
}

\myparab{Baselines.}
We compare \sysname against two baselines: (1) \emph{EPS}: a static electrical-packet-switched network where all links that \sysname could form across any circuit configuration are always active, giving EPS strictly higher total bandwidth; and (2) \emph{Ideal one-shot}: a network reconfigured once before the job starts, with \sysname's total bandwidth divided optimally across parallelism dimensions without port or link granularity constraints, following related work~\cite{wu2025actina}. \rev{The EPS baseline does not model congestion in switches.}

\begin{figure}[h!]
    \centering
    \includegraphics[width=\linewidth]{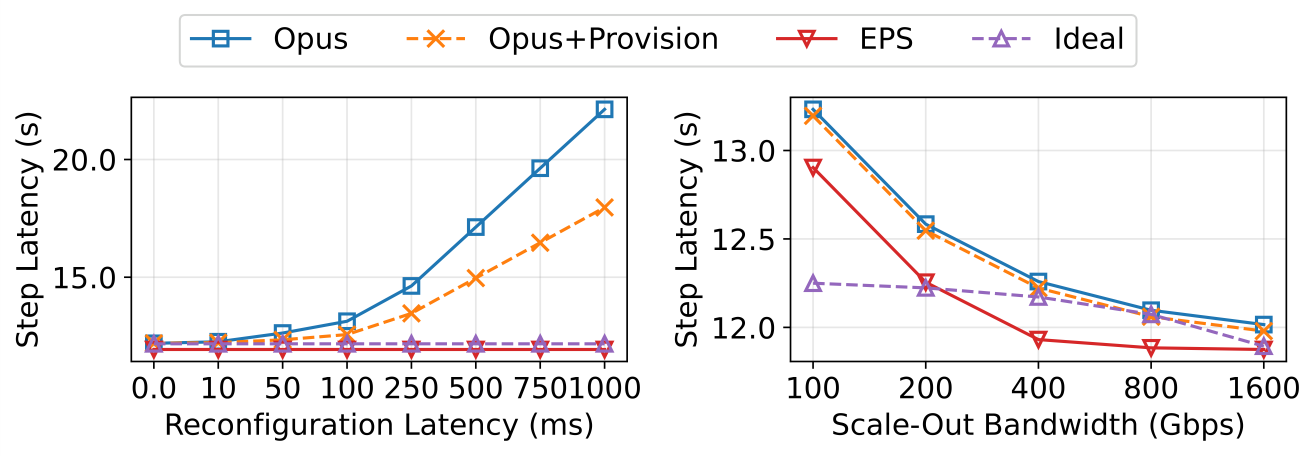}
    \caption{Llama-80B on 128 H200 GPUs (DP=4, PP=4, TP=8). OCS latency sweep at 400\,Gbps scale-out (left); bandwidth sweep at 10\,ms OCS latency (right).}
    \label{fig:sim_llama_1}
\end{figure}

\begin{figure}[h!]
    \centering
    \includegraphics[width=\linewidth]{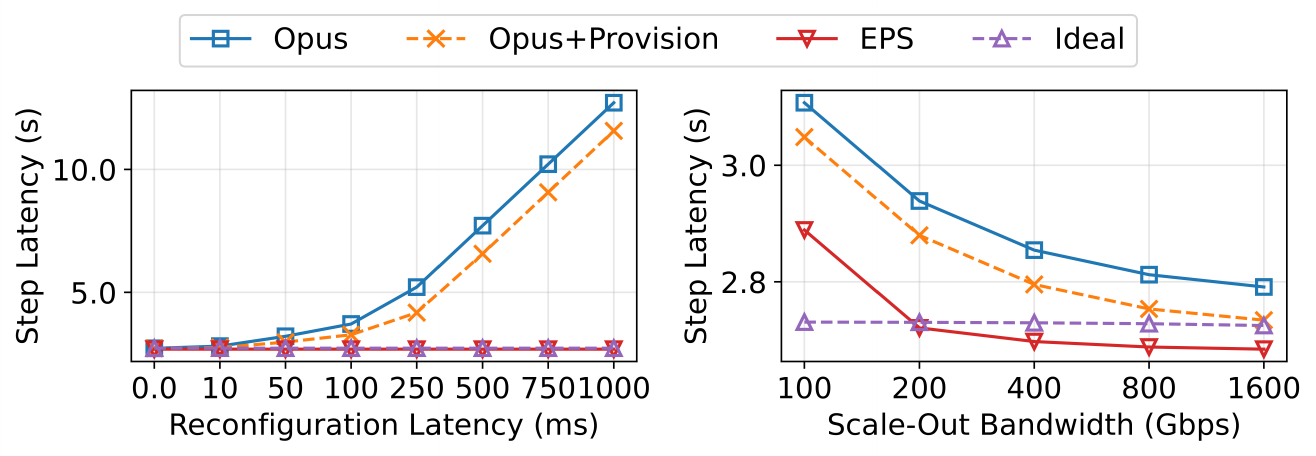}
    \caption{GPT-80B on 512 B200 GPUs (DP=4, PP=4, TP=32). OCS latency sweep at 800\,Gbps scale-out (left); bandwidth sweep at 10\,ms OCS latency (right).}
    \label{fig:sim_gpt_1}
\end{figure}

\myparab{Effect of OCS reconfiguration latency.}
We sweep OCS reconfiguration latency from 0 to 1,000\,ms for two setups: Llama-80B on 128 H200 GPUs (DP=4, PP=4, TP=8, 400\,Gbps scale-out) and GPT-80B on 512 B200 GPUs (DP=4, PP=4, TP=32, 800\,Gbps scale-out), shown in Figures~\ref{fig:sim_llama_1} and~\ref{fig:sim_gpt_1}. \sysname's overhead remains low at production-relevant OCS latencies. On the H200 cluster at 100\,ms OCS latency, \sysname with provisioning is 5.31\% slower than EPS and 3.32\% slower than ideal one-shot. On the B200 cluster at 10\,ms latency, the gaps narrow to 2.49\% and 0.93\% respectively. The GPT model on B200 is more sensitive to reconfiguration latency due to its higher communication intensity.

\begin{figure}[t]
    \centering
    \includegraphics[width=\linewidth]{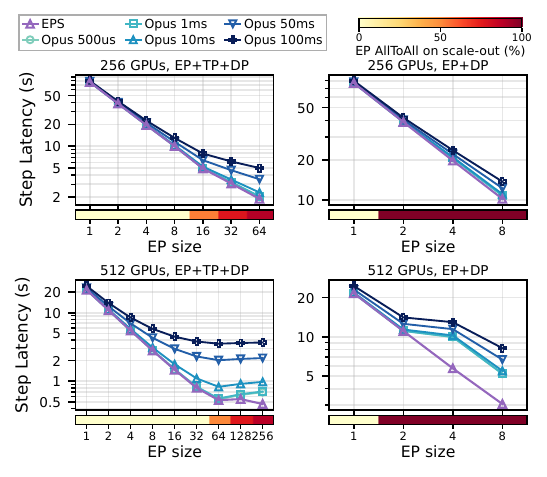}
    \caption{
        \rev{Step latency vs.\ EP degree for \sysname and the EPS baseline. Left: EP co-located with TP/DP. Right: EP spread over DP only, so all EP
    \alltoall crosses scale-out.}
    }
    \label{fig:sim_ep}
\end{figure}

\begin{figure}[h]
    \centering
    \includegraphics[width=\linewidth]{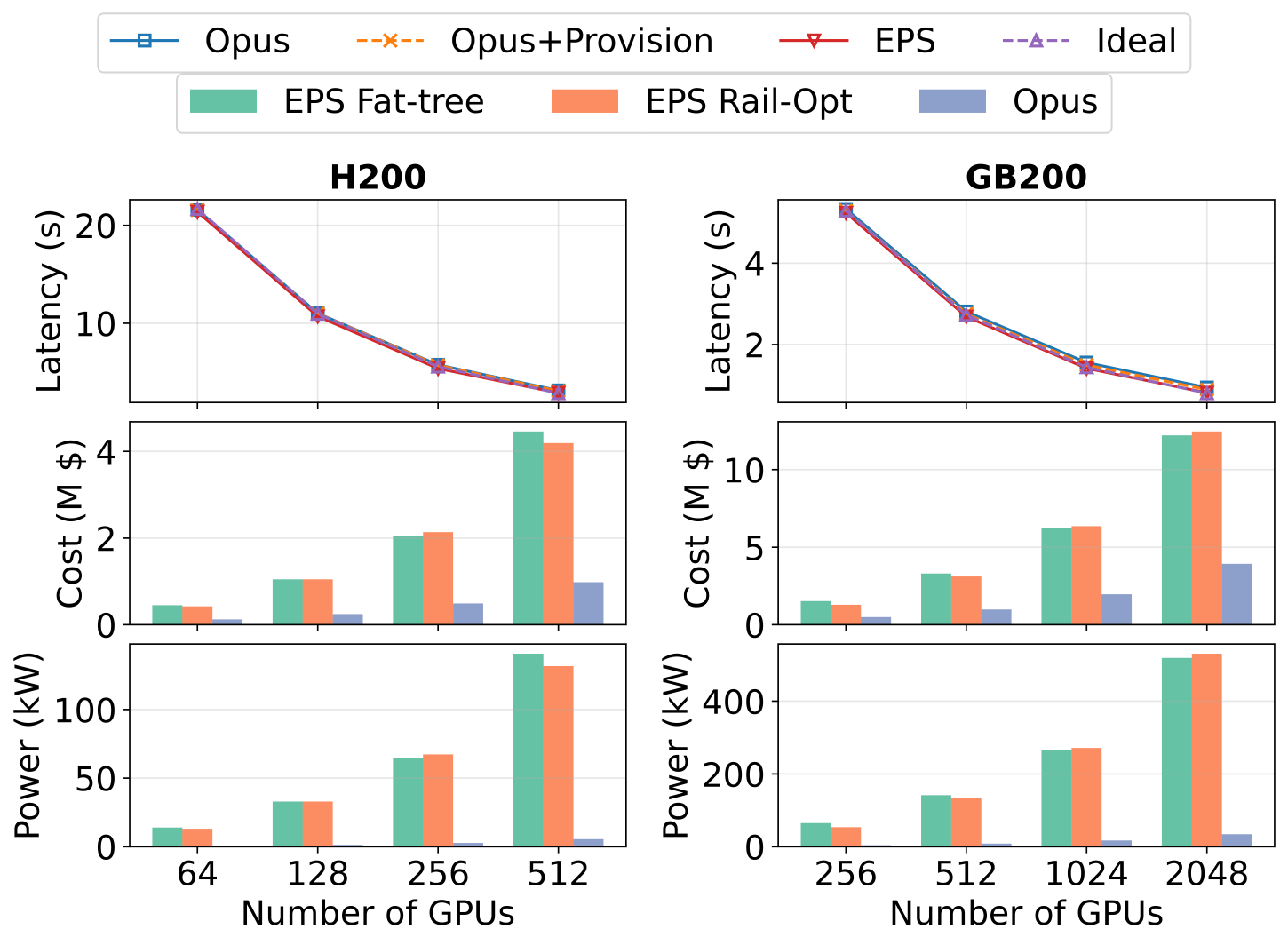}
    \caption{Performance, cost, and power scaling. DGX H200 with 400\,Gbps links (left); B200 with 800\,Gbps links (right). Cost and power exclude fiber cables~\cite{fs_400gbase_sr4_osfp,fs_n9510_64d,800g_tx,800g_switch_power,800g_switch_price,OCS_6000N}.}
    \label{fig:sim_dp_sweep}
\end{figure}

\myparab{Effect of scale-out bandwidth.}
We sweep scale-out bandwidth from 100\,Gbps to 1,600\,Gbps at 10\,ms OCS latency (Figures~\ref{fig:sim_llama_1},~\ref{fig:sim_gpt_1}, right panels). Higher bandwidth reduces \sysname's overhead by shrinking the communication time relative to the fixed reconfiguration cost. Compared to ideal one-shot, \sysname's overhead drops from 7.73\% to 0.72\% on H200 and from 11.63\% to 0.34\% on B200 as bandwidth increases from 100\,Gbps to 1,600\,Gbps. We scale GPU count from 64 to 2,048 by increasing the DP degree while holding TP and PP fixed (Figure~\ref{fig:sim_dp_sweep}). \sysname's overhead remains stable: at 512 H200 GPUs (10 ms OCS, 400\,Gbps), \sysname with provisioning is 6.62\% slower than EPS. At 2,048 B200 GPUs (10 ms OCS, 800 Gbps), the gap is 11.22\%.

\rev{\myparab{Expert-parallelism support.}
\label{sec:sim-ep}
We compare \sysname against EPS by sweeping the fraction of expert-parallel (EP) \alltoall traffic that
crosses the scale-out fabric from 0\% to 100\% (Figure~\ref{fig:sim_ep}). 
We simulate DeepSeek-V3-style MoE model training using Chakra ET trace format: 256 routed experts
with top-8 gating, 
sequence length 4096, and a global batch of 256. 
We use two clusters: a 256$\times$H200 cluster (8$\times$GPU 450\,GB/s scale-up, 400\,Gbps scale-out, TP$=$8, PP$=$4, DP$=$8)
and a 512$\times$H200 cluster (hypothetical 32$\times$GPU 900\,GB/s scale-up, 800\,Gbps
scale-out, TP$=$32, PP$=$2, DP$=$8).
To fully test \sysname performance, we select two expert placement strategies: \emph{EP$+$TP$+$DP}, where EP overlaps with TP (small EP groups stay inside
a scale-up domain) and spills onto scale-out over DP as EP grows~\cite{liu2025moe,yan2026scalable}, and \emph{EP$+$DP}, an adversarial case where EP only overlaps with DP groups~\cite{rajbhandari2022deepspeedmoe}, and 
the EP \alltoall is carried entirely by the scale-out fabric ($100\%$). We scale the EP size from the minimal (EP$=$1) to the maximal (bounded by the pipeline stage size). 

When the EP \alltoall is fully contained by a scale-up domain (0\% on scale-out), \sysname, with provisioning,
matches EPS to within 0.5\% with 0.5\,ms--1\,ms OCS latencies
and within 3\% at 10\,ms. As the EP degree grows and a larger share of the
\alltoall spills onto scale-out, \sysname reconfigures to serve it, and the cost of using multi-hop paths appears (Appendix~\S\ref{sec:alltoall-topo}). With placement EP$+$TP$+$DP at
256 GPUs, \sysname stays within 7.3\% of EPS when 88.9\% of the EP \alltoall
crosses scale-out (0.5\,ms switching), rising to 8.1\% at 1\,ms and 22.5\% at 10\,ms.
The degradation grows with both the scale-out EP fraction and the OCS latency, and is more significant in the adversarial case (EP + DP). 
\sysname performance degradation is bounded because today's large-scale MoE training rarely uses an EP size more than 64~\cite{liu2024deepseek}, and usually EP is overlapped with TP~\cite{liu2025moe,yan2026scalable}.  
}

\subsection{Cost and Power}
\label{sec:cost-power}

\rev{
We estimate the networking cost and power of \sysname and the two EPS baselines, \emph{fat-tree} and \emph{rail-optimized},
with a per-port accounting model. We consider two clusters, a 400G H200 cluster and an 800G B200 cluster. A cluster has $S$ scale-up domains,
each with $G$ GPUs,
for $SG$ GPUs in total, where every GPU drives one scale-out
NIC port through one optical transceiver. The scale-out fabric model contains three devices---electrical switch ports, OCS ports, and transceivers---each with a
fixed unit cost and power (Table~\ref{tab:cost-power-params}). The total cost (and,
identically, the total power) of a topology is
\[
\textstyle
\mathrm{Cost} \,=\, n_{\mathrm{sw}}\,c_{\mathrm{sw}} \,+\, n_{\mathrm{ocs}}\,c_{\mathrm{ocs}} \,+\, n_{\mathrm{xcvr}}\,c_{\mathrm{xcvr}},
\]
where $n_{\mathrm{sw}}$, $n_{\mathrm{ocs}}$, and $n_{\mathrm{xcvr}}$ are the port
counts and $c_\bullet$ are the corresponding unit values.}

\begin{table}[t]
    \centering
    \small
    \setlength{\tabcolsep}{4pt}
    \begin{tabular}{llrr}
        \hline
        \textbf{Component} & \textbf{Config} & \textbf{Power\,(W)} & \textbf{Cost\,(\$)}\\
        \hline
        \makecell[l]{Electrical switch port\\(Tomahawk-4)} & H200, $400$G & $39.4$~\cite{fs_n9510_64d} & $583$~\cite{fs_n9510_64d} \\
        \makecell[l]{Electrical switch port\\(Quantum-X800 XDR)} & B200, $800$G & $48.6$~\cite{800g_switch_power} & $1{,}041$~\cite{800g_switch_price} \\
        \makecell[l]{OCS port\\(Polatis 6000N)} & 400G/800G/1.6T & $0.39$~\cite{OCS_6000N} & $520$~\cite{OCS_6000N} \\
        \makecell[l]{Optical transceiver\\(MMA4Z00-NS)} & $400$G & $10$~\cite{fs_400gbase_sr4_osfp} & $719$~\cite{fs_400gbase_sr4_osfp} \\
        \makecell[l]{Optical transceiver\\(MMA4Z00-NS)} & $800$G & $16$~\cite{800g_tx} & $879$~\cite{800g_tx} \\
        \makecell[l]{Optical transceiver\\(MMS4C11)} & $1.6$T & $30$~\cite{1600g_tx} & $2{,}099$~\cite{1600g_tx} \\
        \hline
    \end{tabular}
    \caption{\rev{Per-port cost and power used in calculation.
    Switch and OCS per-port figures are the device totals divided by port count
    ($64$ ports for the H200 switch, $144$ for the B200 switch, $192$ for the OCS).
    H200 clusters use the $400$G transceiver, B200 clusters the $800$G transceiver,
    and the OCS can be paired with $1.6$T transceivers.}}
    \label{tab:cost-power-params}
\end{table}

\rev{
The topologies differ in their switch, OCS, and transceiver counts. Every GPU NIC
needs one transceiver, so the GPU-side count is $SG$ in all cases. With pluggable
optics each electrical switch port also terminates in a transceiver, so a switched
fabric needs $n_{\mathrm{xcvr}} = SG + n_{\mathrm{sw}}$ transceivers, which grows
with its switch count. Co-packaged-optics (CPO) switches instead integrate the
optics into the switch ASIC and add no separate transceiver, leaving
$n_{\mathrm{xcvr}} = SG$. The H200 baseline uses electrical switches with pluggable optics, and the B200
baseline uses CPO switches~\cite{broadcom_bcm78909}.
\emph{EPS fat-tree} is a three-tier folded-Clos network. With switch radix
$k = \lfloor (4SG)^{1/3} + 0.5 \rfloor$ it uses $k^2 + k^2/4$ switches
($n_{\mathrm{sw}} = k\,(k^2 + k^2/4)$ ports).
\emph{EPS rail-optimized} builds a leaf--spine fat-tree per rail and has a
cluster core layer that interconnects the rails. With switch radix
$k = \lfloor (4S)^{1/2} + 0.5 \rfloor$, it uses
$2G\left\lfloor S/ (k/2) \right\rfloor + \lfloor GS/k \rfloor$
switches ($2\left\lfloor S/ (k/2) \right\rfloor$ leaf--spine
switches per rail, replicated across the $G$ rails, plus $\lfloor GS/k \rfloor$
cluster-core switches). This gives
$n_{\mathrm{sw}} =
k\,(2G\left\lfloor S/ (k/2)  \right\rfloor + \lfloor GS/k \rfloor)$
ports.
\emph{\sysname} replaces the entire electrical fabric with one OCS per rail, so there are $G$ OCSes of radix $2S$, assuming the OCS ports are uni-directional, and one GPU port (bi-directional) is connected to two OCS ports.
This gives $n_{\mathrm{sw}} = 0$, $n_{\mathrm{ocs}} = 2SG$,
and $n_{\mathrm{xcvr}} = SG$, as the OCS is optically transparent and adds no
transceiver.}

\rev{
Figure~\ref{fig:sim_dp_sweep} shows the savings of \sysname based on this cost model in addition to the workload performance.
An OCS port draws about two orders of
magnitude less power than an electrical switch port ($0.39$\,W vs.\ $48.6$\,W at 800G), and
\sysname removes the entire multi-tier switch fabric, so its residual power comes
almost entirely from the GPU transceivers. This yields large power savings of
$15.4\times$ on B200 (rail-optimized, 512--2,048 GPUs) and $23.9\times$ on H200 (rail-optimized, 64--512 GPUs) at a small performance degradation. The cost savings are smaller,
$3.2\times$ and $4.3\times$, because those same GPU transceivers (\$$879$ each at 800G)
remain a large part of \sysname's cost. \sysname saves more compared to H200 than B200 on both metrics
because the H200 pluggable-optics cluster requires a transceiver on every switch port,
all of which \sysname removes, whereas the B200 CPO cluster carries no switch-side transceiver. As OCSes are bandwidth-agnostic, Opus can scale to 1.6T transceivers~\cite{1600g_tx}, matching B200-800G EPS performance (0.2\% slower, Figure~\ref{fig:sim_gpt_1}) at 1.94$\times$ lower network cost and 8.42$\times$ lower power consumption.
}

 \section{Related Work}

\myparab{Optical network designs for general datacenters.}
Prior work has explored optically reconfigurable network topologies using OCSes in traditional datacenter networks, following either traffic-agnostic~\cite{mellette2017rotornet,amir2024shale,shoal,ballani2020sirius} or traffic-aware~\cite{cthrough,farrington2010helios,firefly,projector,chen2017enabling,liang2024negotiator} reconfiguration schemes. Traffic-agnostic designs cycle through topologies at fixed intervals, while traffic-aware designs reconfigure based on estimated demand. These schemes generally target one of two cases: microsecond-scale reconfiguration with shallow buffers for latency-sensitive traffic, or second-scale reconfiguration with deep buffers for elephant flows. Neither regime fits distributed ML workloads, which require both low-latency connectivity and sustained high bandwidth across hundreds of GPUs in a collective communication group. This is the gap \sysname addresses.

\myparab{Reconfigurable optical networks for ML.}
A separate line of work reconfigures optical topologies before a job starts, exploiting ML traffic predictability to allocate bandwidth across parallelism dimensions~\cite{wang2023topoopt,jouppi2023tpu,tpuresilience,morphlux,sipml,liao2025mixnet,wu2024flexible}. However, these designs limit the number of supported parallelisms to the per-accelerator port count---for example, a TPU torus cannot efficiently support more than three parallelism dimensions. Increasing port count strains server architectures or requires technology not yet commercially available~\cite{wu2024flexible}. Recent work on in-job reconfiguration~\cite{wu2025actina} alleviates but does not eliminate this bottleneck.

\myparab{\sysname and rail-optimized fabrics.}
In contrast, \sysname operates within the widely deployed rail-optimized topology~\cite{nvidia-rail-optimize,wang2024rail,gherghescu2024ve,qian2024alibaba} and requires only 2-degree scale-out connectivity to form reconfigurable rings---well within the capabilities of existing server NICs and commercially available OCSes. By time-multiplexing circuits across parallelism phases rather than provisioning static per-dimension links, \sysname decouples the number of supported parallelisms from the per-GPU port count entirely. To our knowledge, \sysname is the first system to demonstrate end-to-end ML training on photonic rails, validated on physical hardware, emulated at medium scale, and simulated at up to 2,048 GPUs.

\section{Discussion}

\myparab{Generality of \sysname.}
\sysname supports up to 10 parallelism dimensions via its topology encoding (\S\ref{sec:design}), easily accommodating the 3--5 dimensions typical of today's large-scale LLM pre-training~\cite{chu2025scaling,liu2024deepseek}. Beyond supervised training, \sysname can serve any ML workload whose communication can be decomposed into predictable phases.
\rev{For instance, OCS reconfiguration schedules can be tied to the KV-cache transfer pipelines in prefill-decode disaggregated inference workloads~\cite{patel2024splitwise,zhong2024distserve}.
} \del{For instance, in MoE models, communication occurs during the routing of tokens to experts (dispatch and combine), which can be profiled similarly to collective operations. The shim can capture these patterns and reconfigure the optical rails accordingly.}Similarly, \sysname can serve RL post-training workloads which alternate between inference and training with distinct parallelism requirements~\cite{wu2025hybridflow}. 
For multi-job environments, \sysname composes per-job topologies into a global OCS configuration, and the non-blocking properties of commodity OCSes~\cite{polatis_series7000} ensure that reconfiguring one job's circuits does not disrupt another's.
\rev{Finally, for \alltoall traffic in expert parallelism, additional scheduling logic can be implemented to provision
optimized topologies through demand prediction \cite{liao2025mixnet}, with its benefit limited by the predictability of the 
future expert \alltoall traffic and the prediction accuracy.}

\myparab{Scalability.}
\sysname scales to large clusters using commodity OCSes. Liquid-crystal OCSes offer up to 512 ports \del{with $\approx$100 ms reconfiguration latency}~\cite{Coherent_OCS_2025}, which suffices to connect large scale-up domains (\eg NVL72) at hyperscaler-relevant scale---up to 18K GPUs per rail. Beyond this, hierarchical \rev{or multi-plane} topologies such as BCube~\cite{sipac,wu2025actina,araujo2026resilient} can extend reach further, though current server and NIC architectures limit the number of per-GPU ports available to support additional hierarchies or network planes. \rev{The ever-increasing switching speed of the optical interconnect (switches and transceivers) can also help address the diminishing window sizes at large scales (\S~\ref{sec:trace})}. 

\myparab{Fault tolerance.}
Photonic rails inherit the fault model of rail-optimized topologies~\cite{wang2024rail} without introducing new failure modes, while reducing failure points by eliminating switch ASICs, transceivers, and lasers from the datapath. Passive optical technologies~\cite{ballani2020sirius,benyahya2025mosaic} and rail redundancy can improve reliability further. \sysname is compatible with existing software-layer fault handling~\cite{gandhi2024recycle,jang2023oobleck,lao2024trainmover} and standard SDN controller replication~\cite{berde2014onos,hunt2010zookeeper}.

\section{Conclusion}
In this work, we propose a novel photonic rail topology for distributed ML training, which offers comparable training performance to state-of-the-art electrical networks but achieves $4\times$ reduction in networking infrastructure cost and over $23\times$ reduction in network power consumption \rev{using existing optical-circuit-switching technologies}. To realize the vision, we develop a complete control plane, \sysname, that orchestrates topology reconfigurations based on the application's communication intents. We deploy the system on a real optical-circuit-switched testbed, evaluate the system performance on the Perlmutter supercomputer, and demonstrate the scalability of its performance and efficiency to 2,000+ GPUs in simulation. 

\textbf{Ethics:} This work does not raise any ethical issues.

\section*{Acknowledgements} We sincerely thank the anonymous SIGCOMM reviewers for their insightful feedback.
This work was supported in part by ACE, one of the seven centers in JUMP 2.0, a Semiconductor Research Corporation (SRC) program sponsored by DARPA. The authors of this work are also supported by NSF Awards No. 2444537 and No. 2435852.
This research used resources of the National Energy Research
Scientific Computing Center, a DOE Office of Science User Facility
supported by the Office of Science of the U.S. Department of Energy
under Contract No. DE-AC02-05CH11231 using NERSC award
ASCR-ERCAP0038386.

\bibliographystyle{ACM-Reference-Format}
\bibliography{reference}

@misc{araujo2026resilient,
      title={Resilient AI Supercomputer Networking using MRC and SRv6}, 
      author={Joao Araujo and Alex Chow and Mark Handley and Ryder Lewis and Christoph Paasch and Jitendra Padhye and Michael Papamichael and Greg Steinbrecher and Amin Tootoonchian and Lihua Yuan and S. Anantharamu and Abhishek Dosi and Mohit Garg and Mahdieh Ghazi and Torsten Hoefler and Deepal Jayasinghe and Jithin Jose and Abdul Kabbani and Guohan Lu and Yang Wang and K. Doddapaneni and Murali Garimella and Vipin Jain and Yanfang Le and H. Nagulapalli and S. Narayanan and Rong Pan and Rathina Sabesan and Raghava Sivaramu and Rip Sohan and Eric Davis and Dragos Dumitrescu and Mohan Kalkunte and Bhaswar Mitra and Guglielmo Morandin and Adrian Popa and Costin Raiciu and Eric Spada and John Spillane and Niranjan Vaidya and Aviv Barnea and Idan Burstein and Elazar Cohen and Yamin Friedman and Noam Katz and Masoud Moshref and Yuval Shpigelman and Shahaf Shuler and Shy Shyman and Sayantan Sur},
      year={2026},
      eprint={2605.04333},
      archivePrefix={arXiv},
      primaryClass={cs.NI},
      url={https://arxiv.org/abs/2605.04333}, 
}

@article{wu2025hybridflow,
  title={HybridFlow: A Flexible and Efficient RLHF Framework},
  author={Wu, Chuan},
  journal={EuroSys 2025 (30/03/2025-03/04/2025, Rotterdam)},
  year={2025}
}

@inproceedings{qian2024alibaba,
author = {Qian, Kun and Xi, Yongqing and Cao, Jiamin and Gao, Jiaqi and Xu, Yichi and Guan, Yu and Fu, Binzhang and Shi, Xuemei and Zhu, Fangbo and Miao, Rui and Wang, Chao and Wang, Peng and Zhang, Pengcheng and Zeng, Xianlong and Ruan, Eddie and Yao, Zhiping and Zhai, Ennan and Cai, Dennis},
title = {Alibaba HPN: A Data Center Network for Large Language Model Training},
year = {2024},
isbn = {9798400706141},
publisher = {Association for Computing Machinery},
address = {New York, NY, USA},
url = {https://doi.org/10.1145/3651890.3672265},
doi = {10.1145/3651890.3672265},
booktitle = {Proceedings of the ACM SIGCOMM 2024 Conference},
pages = {691–706},
numpages = {16},
location = {Sydney, NSW, Australia},
series = {ACM SIGCOMM '24}
}

@manual{ethtool,
  title={ethtool(8) - {Linux} man page},
  author={{Linux}},
  url={https://linux.die.net/man/8/ethtool}
}

@manual{nvidia-mft,
  title={{NVIDIA} Firmware Tools ({MFT}) Documentation},
  author={{NVIDIA}},
  year={2024},
  url={https://docs.nvidia.com/networking/display/nvidia-firmware-tools-mft-documentation-v4-32-0.0.pdf}
}

@inproceedings{wu2025actina,
  title={ACTINA: Adapting Circuit-Switching Techniques for AI Networking Architectures},
  author={Wu, Zhenguo and Klenk, Benjamin and Dennison, Larry and Bergman, Keren},
  booktitle={Proceedings of the International Conference for High Performance Computing, Networking, Storage and Analysis},
  pages={1211--1222},
  year={2025}
}

@inproceedings{benyahya2025mosaic,
author = {Benyahya, Kaoutar and Diaz, Ariel Gomez and Liu, Junyi and Lyutsarev, Vassily and Pantouvaki, Marianna and Shi, Kai and Siew, Shawn Yohanes and Ballani, Hitesh and Burridge, Thomas and Cletheroe, Daniel and Karagiannis, Thomas and Robertson, Brian and Rowstron, Ant and Yang, Mengyang and Costa, Paolo},
title = {Mosaic: Breaking the Optics versus Copper Trade-off with a Wide-and-Slow Architecture and MicroLEDs},
year = {2025},
isbn = {9798400715242},
publisher = {Association for Computing Machinery},
address = {New York, NY, USA},
url = {https://doi.org/10.1145/3718958.3750510},
doi = {10.1145/3718958.3750510},
booktitle = {Proceedings of the ACM SIGCOMM 2025 Conference},
pages = {234–247},
numpages = {14},
location = {S\~{a}o Francisco Convent, Coimbra, Portugal},
series = {SIGCOMM '25}
}

@inproceedings{berde2014onos,
author = {Berde, Pankaj and Gerola, Matteo and Hart, Jonathan and Higuchi, Yuta and Kobayashi, Masayoshi and Koide, Toshio and Lantz, Bob and O'Connor, Brian and Radoslavov, Pavlin and Snow, William and Parulkar, Guru},
title = {ONOS: towards an open, distributed SDN OS},
year = {2014},
isbn = {9781450329897},
publisher = {Association for Computing Machinery},
address = {New York, NY, USA},
url = {https://doi.org/10.1145/2620728.2620744},
doi = {10.1145/2620728.2620744},
booktitle = {Proceedings of the Third Workshop on Hot Topics in Software Defined Networking},
pages = {1–6},
numpages = {6},
location = {Chicago, Illinois, USA},
series = {HotSDN '14}
}

@inproceedings{hunt2010zookeeper,
  title={{ZooKeeper}: Wait-free coordination for internet-scale systems},
  author={Hunt, Patrick and Konar, Mahadev and Junqueira, Flavio P and Reed, Benjamin},
  booktitle={2010 USENIX Annual Technical Conference (USENIX ATC 10)},
  year={2010}
}

@misc{lao2024trainmover,
      title={TrainMover: An Interruption-Resilient Runtime for ML Training}, 
      author={ChonLam Lao and Jiaqi Gao and Jiamin Cao and Zhipeng Zhang and Pengcheng Zhang and Jiangfei Duan and Zhilong Zheng and Yu Guan and Yichi Xu and Yong Li and Zhengping Qian and Aditya Akella and Minlan Yu and Ennan Zhai and Dennis Cai and Jingren Zhou},
      year={2026},
      eprint={2412.12636},
      archivePrefix={arXiv},
      primaryClass={cs.DC},
      url={https://arxiv.org/abs/2412.12636}, 
}

@inproceedings{jang2023oobleck,
  title={Oobleck: Resilient distributed training of large models using pipeline templates},
  author={Jang, Insu and Yang, Zhenning and Zhang, Zhen and Jin, Xin and Chowdhury, Mosharaf},
  booktitle={Proceedings of the 29th Symposium on Operating Systems Principles},
  pages={382--395},
  year={2023}
}

@inproceedings{gandhi2024recycle,
  title={Recycle: Resilient training of large {DNNs} using pipeline adaptation},
  author={Gandhi, Swapnil and Zhao, Mark and Skiadopoulos, Athinagoras and Kozyrakis, Christos},
  booktitle={Proceedings of the ACM SIGOPS 30th Symposium on Operating Systems Principles},
  pages={211--228},
  year={2024}
}

@misc{Coherent_OCS_2025,
  title        = {Optical Circuit Switch (OCS)},
  author       = {{Coherent Corp.}},
  organization = {Coherent Corp.},
  howpublished = {\url{https://www.coherent.com/networking/optical-circuit-switch}},
  year         = {2025},
  note         = {Accessed: 2025-07-10; Based on press release published March 25,2024; Coherent’s liquid‑crystal‑based OCS architecture supports up to 300×300 ports and is optimized for AI/ML data center fabrics},
}

@inproceedings{singla2014high,
  title={High throughput data center topology design},
  author={Singla, Ankit and Godfrey, P Brighten and Kolla, Alexandra},
  booktitle={11th USENIX Symposium on Networked Systems Design and Implementation (NSDI 14)},
  pages={29--41},
  year={2014}
}

@inproceedings{chen2017enabling,
  title={Enabling {Wide-Spread} Communications on Optical Fabric with {MegaSwitch}},
  author={Chen, Li and Chen, Kai and Zhu, Zhonghua and Yu, Minlan and Porter, George and Qiao, Chunming and Zhong, Shan},
  booktitle={14th USENIX Symposium on Networked Systems Design and Implementation (NSDI 17)},
  pages={577--593},
  year={2017}
}

@inproceedings{harsh2020spineless,
  title={Spineless data centers},
  author={Harsh, Vipul and Jyothi, Sangeetha Abdu and Godfrey, P Brighten},
  booktitle={Proceedings of the 19th ACM Workshop on Hot Topics in Networks},
  pages={67--73},
  year={2020}
}

@misc{hpe_cray_ex,
  title        = {HPE Cray EX Supercomputer Overview},
  author       = {{Hewlett Packard Enterprise}},
  year         = {2021},
  howpublished = {\url{https://www.hpe.com/psnow/doc/a50002546enw}},
  note         = {Accessed: 2025-07-09}
}

@inproceedings{gherghescu2024ve,
  title={I've Got 99 Problems But FLOPS Ain't One},
  author={Gherghescu, Alexandru M and B{\u{a}}doiu, Vlad-Andrei and Agache, Alexandru and Dumitru, Mihai-Valentin and Vasilescu, Iuliu and Mantu, Radu and Raiciu, Costin},
  booktitle={Proceedings of the 23rd ACM Workshop on Hot Topics in Networks},
  pages={195--204},
  year={2024}
}

@online{Yeluri2023_power_consumption,
  author       = {Sharada Yeluri},
  title        = {Optimizing Power Consumption in High-End Routers},
  year         = {2023},
  month        = {June},
  day          = {22},
  institution  = {Juniper Networks Elevate Community},
  howpublished = {\url{https://www.linkedin.com/posts/sharada-yeluri_sharadayeluriblogs-junipernetworks-routers-activity}},
}

@inproceedings{agarwal2024harmony,
  title={Harmony: A congestion-free datacenter architecture},
  author={Agarwal, Saksham and Cai, Qizhe and Agarwal, Rachit and Shmoys, David and Vahdat, Amin},
  booktitle={21st USENIX Symposium on Networked Systems Design and Implementation (NSDI 24)},
  pages={329--343},
  year={2024}
}

@article{fedus2022switch,
  title={Switch transformers: Scaling to trillion parameter models with simple and efficient sparsity},
  author={Fedus, William and Zoph, Barret and Shazeer, Noam},
  journal={Journal of Machine Learning Research},
  volume={23},
  number={120},
  pages={1--39},
  year={2022}
}

@article{liu2023ring,
  title={Ring attention with blockwise transformers for near-infinite context},
  author={Liu, Hao and Zaharia, Matei and Abbeel, Pieter},
  journal={arXiv preprint arXiv:2310.01889},
  year={2023}
}

@article{korthikanti2023reducing,
  title={Reducing activation recomputation in large transformer models},
  author={Korthikanti, Vijay Anand and Casper, Jared and Lym, Sangkug and McAfee, Lawrence and Andersch, Michael and Shoeybi, Mohammad and Catanzaro, Bryan},
  journal={Proceedings of Machine Learning and Systems},
  volume={5},
  pages={341--353},
  year={2023}
}

@misc{zhao2023pytorch,
      title={PyTorch FSDP: Experiences on Scaling Fully Sharded Data Parallel}, 
      author={Yanli Zhao and Andrew Gu and Rohan Varma and Liang Luo and Chien-Chin Huang and Min Xu and Less Wright and Hamid Shojanazeri and Myle Ott and Sam Shleifer and Alban Desmaison and Can Balioglu and Pritam Damania and Bernard Nguyen and Geeta Chauhan and Yuchen Hao and Ajit Mathews and Shen Li},
      year={2023},
      eprint={2304.11277},
      archivePrefix={arXiv},
      primaryClass={cs.DC},
      url={https://arxiv.org/abs/2304.11277}, 
}

@article{shoeybi2019megatron,
  title={{Megatron-LM}: Training multi-billion parameter language models using model parallelism},
  author={Shoeybi, Mohammad and Patwary, Mostofa and Puri, Raul and LeGresley, Patrick and Casper, Jared and Catanzaro, Bryan},
  journal={arXiv preprint arXiv:1909.08053},
  year={2019}
}

@inproceedings{rasley2020deepspeed,
  title={{DeepSpeed}: System optimizations enable training deep learning models with over 100 billion parameters},
  author={Rasley, Jeff and Rajbhandari, Samyam and Ruwase, Olatunji and He, Yuxiong},
  booktitle={Proceedings of the 26th ACM SIGKDD international conference on knowledge discovery \& data mining},
  pages={3505--3506},
  year={2020}
}

@misc{liu2025moe,
      title={MoE Parallel Folding: Heterogeneous Parallelism Mappings for Efficient Large-Scale MoE Model Training with Megatron Core}, 
      author={Dennis Liu and Zijie Yan and Xin Yao and Tong Liu and Vijay Korthikanti and Evan Wu and Shiqing Fan and Gao Deng and Hongxiao Bai and Jianbin Chang and Ashwath Aithal and Michael Andersch and Mohammad Shoeybi and Jiajie Yao and Chandler Zhou and David Wu and Xipeng Li and June Yang},
      year={2026},
      eprint={2504.14960},
      archivePrefix={arXiv},
      primaryClass={cs.LG},
      url={https://arxiv.org/abs/2504.14960}, 
}

@misc{yan2026scalable,
      title={Scalable Training of Mixture-of-Experts Models with Megatron Core}, 
      author={Zijie Yan and Hongxiao Bai and Xin Yao and Dennis Liu and Tong Liu and Hongbin Liu and Pingtian Li and Evan Wu and Shiqing Fan and Li Tao and Robin Zhang and Yuzhong Wang and Shifang Xu and Jack Chang and Xuwen Chen and Kunlun Li and Yan Bai and Gao Deng and Nan Zheng and Vijay Anand Korthikanti and Abhinav Khattar and Ethan He and Soham Govande and Sangkug Lym and Zhongbo Zhu and Qi Zhang and Haochen Yuan and Xiaowei Ren and Deyu Fu and Tailai Ma and Shunkang Zhang and Jiang Shao and Ray Wang and Vasudevan Rengasamy and Rachit Garg and Santosh Bhavani and Xipeng Li and Chandler Zhou and David Wu and Yingcan Wei and Ashwath Aithal and Michael Andersch and Mohammad Shoeybi and Jiajie Yao and June Yang},
      year={2026},
      eprint={2603.07685},
      archivePrefix={arXiv},
      primaryClass={cs.DC},
      url={https://arxiv.org/abs/2603.07685}, 
}

@article{man2025stage,
  title={{STAGE}: A Symbolic Tensor grAph GEnerator for distributed {AI} system co-design},
  author={Man, Changhai and Park, Joongun and Wu, Hanjiang and Xu, Huan and Sridharan, Srinivas and Krishna, Tushar},
  journal={arXiv preprint arXiv:2511.10480},
  year={2025}
}

@inproceedings{rajbhandari2022deepspeedmoe,
  title        = {{DeepSpeed-MoE}: Advancing Mixture-of-Experts Inference and Training to Power Next-Generation {AI} Scale},
  author       = {Rajbhandari, Samyam and Li, Conglong and Yao, Zhewei and Zhang, Minjia and Aminabadi, Reza Yazdani and Awan, Ammar Ahmad and Rasley, Jeff and He, Yuxiong},
  booktitle    = {Proceedings of the 39th International Conference on Machine Learning (ICML)},
  year         = {2022}
}

@article{sanders2009two,
  title={Two-tree algorithms for full bandwidth broadcast, reduction and scan},
  author={Sanders, Peter and Speck, Jochen and Tr{\"a}ff, Jesper Larsson},
  journal={Parallel Computing},
  volume={35},
  number={12},
  pages={581--594},
  year={2009},
  publisher={Elsevier}
}

@inproceedings{thakur2003improving,
  title={Improving the performance of collective operations in MPICH},
  author={Thakur, Rajeev and Gropp, William D},
  booktitle={European Parallel Virtual Machine/Message Passing Interface Users’ Group Meeting},
  pages={257--267},
  year={2003},
  organization={Springer}
}

@misc{nvidia2025llama31_405b_dgxc,
  title        = {{Llama-3.1-405B DGXC Benchmarking Recipe}},
  author       = {{NVIDIA}},
  howpublished = {\url{https://catalog.ngc.nvidia.com/orgs/nvidia/teams/dgxc-benchmarking/resources/llama31-405b-dgxc-benchmarking-a}},
  note         = {Version 24.11.1, modified January 29, 2025},
  year         = {2025},
}

@techreport{nvidia2025dgx_h200,
  title        = {NVIDIA DGX H200 Datasheet},
  author       = {{NVIDIA Corporation}},
  institution  = {NVIDIA Corporation},
  address      = {Santa Clara, CA},
  year         = {2025},
  month        = jul,
  type         = {Datasheet},
  url          = {https://resources.nvidia.com/en-us-dgx-systems/dgx-h200-datasheet},
  note         = {Includes specifications of the DGX H200 system, featuring 8× H200 GPUs, dual Xeon Platinum 8480C CPUs, 2 TB system memory, 30 TB NVMe SSD, and full NVIDIA AI Enterprise software stack}  
}

@manual{nvidia_nccl_communicators,
  title        = "{NVIDIA Collective Communication Library (NCCL)}: Creating a Communicator",
  author       = "{NVIDIA Corporation}",
  organization = "{NVIDIA}",
  year         = "2020",
  url          = "https://docs.nvidia.com/deeplearning/nccl/user-guide/docs/usage/communicators.html",
  note         = "Accessed July 6, 2025",
}

@misc{broadcom_bcm78909,
  author       = {{Broadcom Inc.}},
  title        = {BCM78909 51.2‑Tb/s Multilayer Co‑Packaged Optics Switch},
  year         = {2025},
  howpublished = {Online; accessed July 5, 2025},
  url          = {https://www.broadcom.com/products/fiber-optic-modules-components/co-packaged-optics/switches/bcm78909},
  note         = {A high‑radix, high‑bandwidth CPO switch supporting up to 64×800GbE or 128×400GbE.}
}

@misc{nvidia_silicon_photonics,
  author       = {{NVIDIA Corporation}},
  title        = {Co‑Packaged Silicon Photonics Networking Switches},
  year         = {2025},
  howpublished = {Online; accessed July 5,2025},
  url          = {https://www.nvidia.com/en-us/networking/products/silicon-photonics/},
  note         = {Describes NVIDIA’s co‑packaged optics (CPO) switches with integrated silicon photonics}  
}

@misc{nersc_perlmutter_architecture,
  author       = "{National Energy Research Scientific Computing Center (NERSC)}",
  title        = "{Perlmutter Architecture — NERSC Documentation}",
  howpublished = "\url{https://docs.nersc.gov/systems/perlmutter/architecture/}",
  year         = "2025",
  note         = "Accessed: 2025-07-04",
}

@misc{broadcom_cpo,
  author       = {{Broadcom Inc.}},
  title        = {Co‑Packaged Optics (CPO)},
  howpublished = {\url{https://www.broadcom.com/info/optics/cpo}},
  year         = {2025},
  note         = {Accessed: 2025-07-03},
  publisher    = {Broadcom Inc.},
  address      = {Irving, TX},
}

@misc{fs_n9510_64d,
  author       = {{FS.COM}},
  title        = {{N9510‑64D 64‑Port Ethernet L3 Data Center Switch (Broadcom Tomahawk‑4, 64×400GbE)}},
  howpublished = {\url{https://www.fs.com/products/149853.html}},
  year         = {n.d.},
  note         = {Accessed: 2025-07-02},
}

@misc{nvidia_connectx7_datasheet,
  author       = {{NVIDIA Corporation}},
  title        = {{ConnectX‑7 400G Adapters Datasheet}},
  howpublished = {\url{https://resources.nvidia.com/en-us-accelerated-networking-resource-library/connectx-7-datasheet}},
  year         = {2024},
  note         = {Accessed: 2025-07-02},
}

@misc{fs_400gbase_sr4_osfp,
  author       = {{FS.COM}},
  title        = {{NVIDIA/Mellanox MMA4Z00-NS400 Compatible 400GBASE-SR4 OSFP RHS/Flat Top 4$\times$100G PAM4 850nm 50m MPO-12/APC MMF Ethernet Transceiver Module}},
  howpublished = {\url{https://www.fs.com/products/310009.html?now_cid=3652}},
  year         = {n.d.},
  note         = {Accessed: 2026-06-20},
}

@misc{polatis_series7000,
  author       = {{Polatis (a HUBER+SUHNER company)}},
  title        = {Series 7000 - 384x384-port Software-Defined Optical Circuit Switch},
  howpublished = {\url{https://www.polatis.com/series-7000-384x384-port-software-controlled-optical-circuit-switch-sdn-enabled.asp}},
  year         = {n.d.},
  note         = {Accessed: 2025-07-01}
}

@inproceedings{ballani2020sirius,
author = {Ballani, Hitesh and Costa, Paolo and Behrendt, Raphael and Cletheroe, Daniel and Haller, Istvan and Jozwik, Krzysztof and Karinou, Fotini and Lange, Sophie and Shi, Kai and Thomsen, Benn and Williams, Hugh},
title = {Sirius: A Flat Datacenter Network with Nanosecond Optical Switching},
year = {2020},
isbn = {9781450379557},
publisher = {Association for Computing Machinery},
address = {New York, NY, USA},
url = {https://doi.org/10.1145/3387514.3406221},
doi = {10.1145/3387514.3406221},
booktitle = {Proceedings of the Annual Conference of the ACM Special Interest Group on Data Communication on the Applications, Technologies, Architectures, and Protocols for Computer Communication},
pages = {782–797},
numpages = {16},
location = {Virtual Event, USA},
series = {SIGCOMM '20}
}

@inproceedings{liang2024negotiator,
  title={NegotiaToR: Towards A Simple Yet Effective On-demand Reconfigurable Datacenter Network},
  author={Liang, Cong and Song, Xiangli and Cheng, Jing and Wang, Mowei and Liu, Yashe and Liu, Zhenhua and Zhao, Shizhen and Cui, Yong},
  booktitle={Proceedings of the ACM SIGCOMM 2024 Conference},
  pages={415--432},
  year={2024}
}

@inproceedings{amir2024shale,
  title={Shale: A practical, scalable oblivious reconfigurable network},
  author={Amir, Daniel and Saran, Nitika and Wilson, Tegan and Kleinberg, Robert and Shrivastav, Vishal and Weatherspoon, Hakim},
  booktitle={Proceedings of the ACM SIGCOMM 2024 Conference},
  pages={449--464},
  year={2024}
}

@inproceedings{liao2025mixnet,
author = {Liao, Xudong and Sun, Yijun and Tian, Han and Wan, Xinchen and Jin, Yilun and Wang, Zilong and Ren, Zhenghang and Huang, Xinyang and Li, Wenxue and Tse, Kin Fai and Zhong, Zhizhen and Liu, Guyue and Zhang, Ying and Ye, Xiaofeng and Zhang, Yiming and Chen, Kai},
title = {MixNet: A Runtime Reconfigurable Optical-Electrical Fabric for Distributed Mixture-of-Experts Training},
year = {2025},
isbn = {9798400715242},
publisher = {Association for Computing Machinery},
address = {New York, NY, USA},
url = {https://doi.org/10.1145/3718958.3750465},
doi = {10.1145/3718958.3750465},
booktitle = {Proceedings of the ACM SIGCOMM 2025 Conference},
pages = {554–574},
numpages = {21},
location = {S\~{a}o Francisco Convent, Coimbra, Portugal},
series = {SIGCOMM '25}
}

@manual{nvidia_hgx_platform_2025,
  title        = {{NVIDIA HGX Platform}},
  author       = {{NVIDIA Corporation}},
  organization = {NVIDIA},
  year         = {2025},
  note         = {Reference architecture combining GPUs, NVLink/NVSwitch, networking, and AI/HPC software stack},
  url          = {https://www.nvidia.com/en-us/data-center/hgx/}
}

@manual{nvidia_dgx_superpod_2025,
  title        = {{NVIDIA DGX SuperPOD}},
  author       = {{NVIDIA Corporation}},
  organization = {NVIDIA},
  year         = {2025},
  month        = jun,
  note         = {Full-stack data center platform scaling to tens of thousands of GPUs; includes compute, networking, storage, and software} ,
  url          = {https://www.nvidia.com/en-us/data-center/dgx-superpod/}
}

@inproceedings{gangidi2024rdma,
author = {Gangidi, Adithya and Miao, Rui and Zheng, Shengbao and Bondu, Sai Jayesh and Goes, Guilherme and Morsy, Hany and Puri, Rohit and Riftadi, Mohammad and Shetty, Ashmitha Jeevaraj and Yang, Jingyi and Zhang, Shuqiang and Fernandez, Mikel Jimenez and Gandham, Shashidhar and Zeng, Hongyi},
title = {RDMA over Ethernet for Distributed Training at Meta Scale},
year = {2024},
isbn = {9798400706141},
publisher = {Association for Computing Machinery},
address = {New York, NY, USA},
url = {https://doi.org/10.1145/3651890.3672233},
doi = {10.1145/3651890.3672233},
booktitle = {Proceedings of the ACM SIGCOMM 2024 Conference},
pages = {57–70},
numpages = {14},
location = {Sydney, NSW, Australia},
series = {ACM SIGCOMM '24}
}

@manual{nvidia_nccl,
  title        = {{NVIDIA Collective Communications Library (NCCL)}},
  author       = {{NVIDIA Corporation}},
  organization = {NVIDIA Developer},
  year         = {2025},
  note         = {Version 2.x; MPI-compatible multi‑GPU / multi‑node collective communication library},
  url          = {https://developer.nvidia.com/nccl}
}

@techreport{nvidia_spectrum-x_2025,
  title       = {NVIDIA Announces Spectrum‑X Photonics, Co‑Packaged Optics Networking Switches to Scale AI Factories to Millions of GPUs},
  author      = {{NVIDIA Corporation}},
  institution = {NVIDIA Corporation},
  type        = {Press Release},
  address     = {Santa Clara, CA, USA},
  month       = mar,
  day         = {18},
  year        = {2025},
  note        = {Unveiled at GTC 2025},
  url         = {https://nvidianews.nvidia.com/news/nvidia-spectrum-x-co-packaged-optics-networking-switches-ai-factories}
}

@techreport{broadcom_scip_2021,
  title        = {SiPh Chiplets In Package (SCIP)},
  author       = {{Broadcom Inc.\, Optical Systems Division}},
  institution  = {Broadcom Inc.},
  type         = {Technical Report},
  address      = {Irvine, CA, USA},
  month        = nov,
  year         = {2021},
  note         = {OSD CPO SCIP\_20211106 V5},
  url          = {https://docs.broadcom.com/doc/siph-chiplets-in-package-scip}
}

@inproceedings{chu2025scaling,
author = {Chu, Weiwei and Xie, Xinfeng and Yu, Jiecao and Wang, Jie and Phanishayee, Amar and Tang, Chunqiang and Hao, Yuchen and Huang, Jianyu and Ozdal, Mustafa and Wang, Jun and Goswami, Vedanuj and Goyal, Naman and Kadian, Abhishek and Gu, Andrew and Cai, Chris and Tian, Feng and Wang, Xiaodong and Si, Min and Balaji, Pavan and Chu, Ching-Hsiang and Park, Jongsoo},
title = {Scaling Llama 3 Training with Efficient Parallelism Strategies},
year = {2025},
isbn = {9798400712616},
publisher = {Association for Computing Machinery},
address = {New York, NY, USA},
url = {https://doi.org/10.1145/3695053.3731410},
doi = {10.1145/3695053.3731410},
booktitle = {Proceedings of the 52nd Annual International Symposium on Computer Architecture},
pages = {1703–1716},
numpages = {14},
location = {
},
series = {ISCA '25}
}

@misc{liang2024torchtitan,
      title={TorchTitan: One-stop PyTorch native solution for production ready LLM pre-training}, 
      author={Wanchao Liang and Tianyu Liu and Less Wright and Will Constable and Andrew Gu and Chien-Chin Huang and Iris Zhang and Wei Feng and Howard Huang and Junjie Wang and Sanket Purandare and Gokul Nadathur and Stratos Idreos},
      year={2025},
      eprint={2410.06511},
      archivePrefix={arXiv},
      primaryClass={cs.CL},
      url={https://arxiv.org/abs/2410.06511}, 
}

@misc{Lumentum2025OCS,
  author       = {{Lumentum Holdings Inc.}},
  title        = {{Lumentum Optical Circuit Switch to Improve Next‑Generation AI Data Center Scalability}},
  howpublished = {\url{https://www.lumentum.com/en/media-room/news-releases/lumentum-optical-circuit-switch-improve-next-generation-ai-data-center}},
  month        = mar,
  year         = {2025},
  day          = {26},
  note         = {Accessed June 20, 2025},
}

@article{qi2023zero,
  title={Zero bubble pipeline parallelism},
  author={Qi, Penghui and Wan, Xinyi and Huang, Guangxing and Lin, Min},
  journal={arXiv preprint arXiv:2401.10241},
  year={2023}
}

@misc{liu2024deepseek,
      title={DeepSeek-V3 Technical Report}, 
      author={DeepSeek-AI and Aixin Liu and Bei Feng and Bing Xue and Bingxuan Wang and Bochao Wu and Chengda Lu and Chenggang Zhao and Chengqi Deng and Chenyu Zhang and Chong Ruan and Damai Dai and Daya Guo and Dejian Yang and Deli Chen and Dongjie Ji and Erhang Li and Fangyun Lin and Fucong Dai and Fuli Luo and Guangbo Hao and Guanting Chen and Guowei Li and H. Zhang and Han Bao and Hanwei Xu and Haocheng Wang and Haowei Zhang and Honghui Ding and Huajian Xin and Huazuo Gao and Hui Li and Hui Qu and J. L. Cai and Jian Liang and Jianzhong Guo and Jiaqi Ni and Jiashi Li and Jiawei Wang and Jin Chen and Jingchang Chen and Jingyang Yuan and Junjie Qiu and Junlong Li and Junxiao Song and Kai Dong and Kai Hu and Kaige Gao and Kang Guan and Kexin Huang and Kuai Yu and Lean Wang and Lecong Zhang and Lei Xu and Leyi Xia and Liang Zhao and Litong Wang and Liyue Zhang and Meng Li and Miaojun Wang and Mingchuan Zhang and Minghua Zhang and Minghui Tang and Mingming Li and Ning Tian and Panpan Huang and Peiyi Wang and Peng Zhang and Qiancheng Wang and Qihao Zhu and Qinyu Chen and Qiushi Du and R. J. Chen and R. L. Jin and Ruiqi Ge and Ruisong Zhang and Ruizhe Pan and Runji Wang and Runxin Xu and Ruoyu Zhang and Ruyi Chen and S. S. Li and Shanghao Lu and Shangyan Zhou and Shanhuang Chen and Shaoqing Wu and Shengfeng Ye and Shengfeng Ye and Shirong Ma and Shiyu Wang and Shuang Zhou and Shuiping Yu and Shunfeng Zhou and Shuting Pan and T. Wang and Tao Yun and Tian Pei and Tianyu Sun and W. L. Xiao and Wangding Zeng and Wanjia Zhao and Wei An and Wen Liu and Wenfeng Liang and Wenjun Gao and Wenqin Yu and Wentao Zhang and X. Q. Li and Xiangyue Jin and Xianzu Wang and Xiao Bi and Xiaodong Liu and Xiaohan Wang and Xiaojin Shen and Xiaokang Chen and Xiaokang Zhang and Xiaosha Chen and Xiaotao Nie and Xiaowen Sun and Xiaoxiang Wang and Xin Cheng and Xin Liu and Xin Xie and Xingchao Liu and Xingkai Yu and Xinnan Song and Xinxia Shan and Xinyi Zhou and Xinyu Yang and Xinyuan Li and Xuecheng Su and Xuheng Lin and Y. K. Li and Y. Q. Wang and Y. X. Wei and Y. X. Zhu and Yang Zhang and Yanhong Xu and Yanhong Xu and Yanping Huang and Yao Li and Yao Zhao and Yaofeng Sun and Yaohui Li and Yaohui Wang and Yi Yu and Yi Zheng and Yichao Zhang and Yifan Shi and Yiliang Xiong and Ying He and Ying Tang and Yishi Piao and Yisong Wang and Yixuan Tan and Yiyang Ma and Yiyuan Liu and Yongqiang Guo and Yu Wu and Yuan Ou and Yuchen Zhu and Yuduan Wang and Yue Gong and Yuheng Zou and Yujia He and Yukun Zha and Yunfan Xiong and Yunxian Ma and Yuting Yan and Yuxiang Luo and Yuxiang You and Yuxuan Liu and Yuyang Zhou and Z. F. Wu and Z. Z. Ren and Zehui Ren and Zhangli Sha and Zhe Fu and Zhean Xu and Zhen Huang and Zhen Zhang and Zhenda Xie and Zhengyan Zhang and Zhewen Hao and Zhibin Gou and Zhicheng Ma and Zhigang Yan and Zhihong Shao and Zhipeng Xu and Zhiyu Wu and Zhongyu Zhang and Zhuoshu Li and Zihui Gu and Zijia Zhu and Zijun Liu and Zilin Li and Ziwei Xie and Ziyang Song and Ziyi Gao and Zizheng Pan},
      year={2025},
      eprint={2412.19437},
      archivePrefix={arXiv},
      primaryClass={cs.CL},
      url={https://arxiv.org/abs/2412.19437}, 
}

@inproceedings{wang2024rail,
  title={Rail-only: A low-cost high-performance network for training LLMs with trillion parameters},
  author={Wang, Weiyang and Ghobadi, Manya and Shakeri, Kayvon and Zhang, Ying and Hasani, Naader},
  booktitle={2024 IEEE Symposium on High-Performance Interconnects (HOTI)},
  pages={1--10},
  year={2024},
  organization={IEEE}
}

@article{wu2024flexible,
  title={Flexible silicon photonic architecture for accelerating distributed deep learning},
  author={Wu, Zhenguo and Yuan Dai, Liang and Wang, Yuyang and Wang, Songli and Bergman, Keren},
  journal={Journal of Optical Communications and Networking},
  volume={16},
  number={2},
  pages={A157--A168},
  year={2024},
  publisher={Optica Publishing Group}
}

@misc{ultrascale_playbook,
  title = {The Ultra-Scale Playbook: Training LLMs on GPU Clusters},
  author = {Nouamane Tazi and Ferdinand Mom and Haojun Zhao and Phuc Nguyen and Mohamed Mekkouri and Leandro Werra and Thomas Wolf},
  year = {2025},
  howpublished = {\url{https://huggingface.co/spaces/nanotron/ultrascale-playbook}},
  note = {Accessed: 2025-05-16}
}

@inproceedings{farrington2010helios,
  title={Helios: a hybrid electrical/optical switch architecture for modular data centers},
  author={Farrington, Nathan and Porter, George and Radhakrishnan, Sivasankar and Bazzaz, Hamid Hajabdolali and Subramanya, Vikram and Fainman, Yeshaiahu and Papen, George and Vahdat, Amin},
  booktitle={Proceedings of the ACM SIGCOMM 2010 Conference},
  pages={339--350},
  year={2010}
}

@inproceedings{mellette2017rotornet,
  title={{RotorNet}: A scalable, low-complexity, optical datacenter network},
  author={Mellette, William M and McGuinness, Rob and Roy, Arjun and Forencich, Alex and Papen, George and Snoeren, Alex C and Porter, George},
  booktitle={Proceedings of the Conference of the ACM Special Interest Group on Data Communication},
  pages={267--280},
  year={2017}
}

@inproceedings{wang2023topoopt,
  title={{TopoOpt}: Co-optimizing network topology and parallelization strategy for distributed training jobs},
  author={Wang, Weiyang and Khazraee, Moein and Zhong, Zhizhen and Ghobadi, Manya and Jia, Zhihao and Mudigere, Dheevatsa and Zhang, Ying and Kewitsch, Anthony},
  booktitle={20th USENIX Symposium on Networked Systems Design and Implementation (NSDI 23)},
  pages={739--767},
  year={2023}
}

@inproceedings{jouppi2023tpu,
author = {Jouppi, Norm and Kurian, George and Li, Sheng and Ma, Peter and Nagarajan, Rahul and Nai, Lifeng and Patil, Nishant and Subramanian, Suvinay and Swing, Andy and Towles, Brian and Young, Clifford and Zhou, Xiang and Zhou, Zongwei and Patterson, David A},
title = {TPU v4: An Optically Reconfigurable Supercomputer for Machine Learning with Hardware Support for Embeddings},
year = {2023},
isbn = {9798400700958},
publisher = {Association for Computing Machinery},
address = {New York, NY, USA},
url = {https://doi.org/10.1145/3579371.3589350},
doi = {10.1145/3579371.3589350},
booktitle = {Proceedings of the 50th Annual International Symposium on Computer Architecture},
articleno = {82},
numpages = {14},
location = {Orlando, FL, USA},
series = {ISCA '23}
}

@inproceedings{tpuresilience,
  author        = {Yazhou Zu and Alireza Ghaffarkhah and Hoang-Vu Dang and Brian Towles and Steven Hand and Safeen Huda and Adekunle Bello and Alexander Kolbasov and Arash Rezaei and Dayou Du and Steve Lacy and Hang Wang and Aaron Wisner and Chris Lewis and Henri Bahini},
  title         = {Resiliency at Scale: Managing {Google{\textquoteright}s} {TPUv4} Machine Learning Supercomputer},
  booktitle     = {21st USENIX Symposium on Networked Systems Design and Implementation (NSDI 24)},
  year          = {2024},
  pages         = {761--774},
  month         = apr
}

@inproceedings{vl2,
  author        = {Greenberg, Albert and Hamilton, James R. and Jain, Navendu and Kandula, Srikanth and Kim, Changhoon and Lahiri, Parantap and Maltz, David A. and Patel, Parveen and Sengupta, Sudipta},
  title         = {VL2: A Scalable and Flexible Data Center Network},
  year          = {2009},
  isbn          = {9781605585949},
  publisher     = {Association for Computing Machinery},
  address       = {New York, NY, USA},
  url           = {https://doi.org/10.1145/1592568.1592576},
  doi           = {10.1145/1592568.1592576},
  booktitle     = {Proceedings of the ACM SIGCOMM 2009 Conference on Data Communication},
  pages         = {51–62},
  numpages      = {12},
  location      = {Barcelona, Spain},
  series        = {SIGCOMM '09}
}

@article{jupiter-rising,
  author        = {Singh, Arjun and Ong, Joon and Agarwal, Amit and Anderson, Glen and Armistead, Ashby and Bannon, Roy and Boving, Seb and Desai, Gaurav and Felderman, Bob and Germano, Paulie and Kanagala, Anand and Provost, Jeff and Simmons, Jason and Tanda, Eiichi and Wanderer, Jim and H\"{o}lzle, Urs and Stuart, Stephen and Vahdat, Amin},
  title         = {Jupiter Rising: A Decade of Clos Topologies and Centralized Control in Google's Datacenter Network},
  year          = {2015},
  issue_date    = {October 2015},
  publisher     = {Association for Computing Machinery},
  address       = {New York, NY, USA},
  volume        = {45},
  number        = {4},
  issn          = {0146-4833},
  url           = {https://doi.org/10.1145/2829988.2787508},
  doi           = {10.1145/2829988.2787508},
  journal       = {SIGCOMM Comput. Commun. Rev.},
  month         = {aug},
  pages         = {183–197},
  numpages      = {15}
}

@inproceedings{jupiter-evolving,
  author        = {Poutievski, Leon and Mashayekhi, Omid and Ong, Joon and Singh, Arjun and Tariq, Mukarram and Wang, Rui and Zhang, Jianan and Beauregard, Virginia and Conner, Patrick and Gribble, Steve and Kapoor, Rishi and Kratzer, Stephen and Li, Nanfang and Liu, Hong and Nagaraj, Karthik and Ornstein, Jason and Sawhney, Samir and Urata, Ryohei and Vicisano, Lorenzo and Yasumura, Kevin and Zhang, Shidong and Zhou, Junlan and Vahdat, Amin},
  title         = {{Jupiter Evolving: Transforming Google's Datacenter Network via Optical Circuit Switches and Software-Defined Networking}},
  year          = {2022},
  booktitle     = {Proceedings of the ACM SIGCOMM 2022 Conference},
  pages         = {66–85},
  numpages      = {20},
  series        = {SIGCOMM '22}
}

@inproceedings{cthrough,
  author        = {Wang, Guohui and Andersen, David G. and Kaminsky, Michael and Papagiannaki, Konstantina and Ng, T.S. Eugene and Kozuch, Michael and Ryan, Michael},
  title         = {C-Through: Part-Time Optics in Data Centers},
  year          = {2010},
  isbn          = {9781450302012},
  publisher     = {Association for Computing Machinery},
  address       = {New York, NY, USA},
  url           = {https://doi.org/10.1145/1851182.1851222},
  doi           = {10.1145/1851182.1851222},
  booktitle     = {Proceedings of the ACM SIGCOMM 2010 Conference},
  pages         = {327–338},
  numpages      = {12},
  location      = {New Delhi, India},
  series        = {SIGCOMM '10}
}

@inproceedings{shoal,
  author        = {Vishal Shrivastav and Asaf Valadarsky and Hitesh Ballani and Paolo Costa and Ki Suh Lee and Han Wang and Rachit Agarwal and Hakim Weatherspoon},
  title         = {Shoal: A Network Architecture for Disaggregated Racks},
  booktitle     = {16th USENIX Symposium on Networked Systems Design and Implementation (NSDI 19)},
  year          = {2019},
  isbn          = {978-1-931971-49-2},
  address       = {Boston, MA},
  pages         = {255--270},
  url           = {https://www.usenix.org/conference/nsdi19/presentation/shrivastav},
  publisher     = {USENIX Association},
  month         = feb
}

@inproceedings{firefly,
  author        = {Hamedazimi, Navid and Qazi, Zafar and Gupta, Himanshu and Sekar, Vyas and Das, Samir R. and Longtin, Jon P. and Shah, Himanshu and Tanwer, Ashish},
  title         = {FireFly: A Reconfigurable Wireless Data Center Fabric Using Free-Space Optics},
  year          = {2014},
  isbn          = {9781450328364},
  publisher     = {Association for Computing Machinery},
  address       = {New York, NY, USA},
  url           = {https://doi.org/10.1145/2619239.2626328},
  doi           = {10.1145/2619239.2626328},
  booktitle     = {Proceedings of the 2014 ACM Conference on SIGCOMM},
  pages         = {319–330},
  numpages      = {12},
  location      = {Chicago, Illinois, USA},
  series        = {SIGCOMM '14}
}

@inproceedings{projector,
  author        = {Ghobadi, Monia and Mahajan, Ratul and Phanishayee, Amar and Devanur, Nikhil and Kulkarni, Janardhan and Ranade, Gireeja and Blanche, Pierre-Alexandre and Rastegarfar, Houman and Glick, Madeleine and Kilper, Daniel},
  title         = {ProjecToR: Agile Reconfigurable Data Center Interconnect},
  year          = {2016},
  isbn          = {9781450341936},
  publisher     = {Association for Computing Machinery},
  address       = {New York, NY, USA},
  url           = {https://doi.org/10.1145/2934872.2934911},
  doi           = {10.1145/2934872.2934911},
  booktitle     = {Proceedings of the 2016 ACM SIGCOMM Conference},
  pages         = {216–229},
  numpages      = {14},
  location      = {Florianopolis, Brazil},
  series        = {SIGCOMM '16}
}

@inproceedings{sipml,
  title={{SiP-ML}: high-bandwidth optical network interconnects for machine learning training},
  author={Khani, Mehrdad and Ghobadi, Manya and Alizadeh, Mohammad and Zhu, Ziyi and Glick, Madeleine and Bergman, Keren and Vahdat, Amin and Klenk, Benjamin and Ebrahimi, Eiman},
  booktitle={Proceedings of the 2021 ACM SIGCOMM 2021 Conference},
  pages={657--675},
  year={2021}
}

@inproceedings{lumorph,
author = {Kumar, Abhishek Vijaya and Devraj, Arjun and Bunandar, Darius and Singh, Rachee},
title = {A case for server-scale photonic connectivity},
year = {2024},
isbn = {9798400712722},
publisher = {Association for Computing Machinery},
address = {New York, NY, USA},
url = {https://doi.org/10.1145/3696348.3696856},
doi = {10.1145/3696348.3696856},
booktitle = {Proceedings of the 23rd ACM Workshop on Hot Topics in Networks},
pages = {290–299},
numpages = {10},
location = {Irvine, CA, USA},
series = {HotNets '24}
}

@inproceedings{sipac,
  author        = {Wu, Zhenguo and Dai, Liang Yuan and Zhu, Ziyi and Novick, Asher and Glick, Madeleine and Bergman, Keren},
  booktitle     = {2023 Optical Fiber Communications Conference and Exhibition (OFC)},
  title         = {{SiP} Architecture for Accelerating Collective Communication in Distributed Deep Learning},
  year          = {2023},
  volume        = {},
  number        = {},
  pages         = {1--3},
  doi           = {10.1364/OFC.2023.W1G.1}
}

@inproceedings{slimfly,
  author        = {Besta, Maciej and Hoefler, Torsten},
  booktitle     = {SC '14: Proceedings of the International Conference for High Performance Computing, Networking, Storage and Analysis},
  title         = {Slim Fly: A Cost Effective Low-Diameter Network Topology},
  year          = {2014},
  volume        = {},
  number        = {},
  pages         = {348--359},
  doi           = {10.1109/SC.2014.34}
}

@inproceedings{jellyfish,
  author        = {Ankit Singla and Chi-Yao Hong and Lucian Popa and P. Brighten Godfrey},
  title         = {Jellyfish: Networking Data Centers Randomly},
  booktitle     = {9th USENIX Symposium on Networked Systems Design and Implementation (NSDI 12)},
  year          = {2012},
  isbn          = {978-931971-92-8},
  address       = {San Jose, CA},
  pages         = {225--238},
  url           = {https://www.usenix.org/conference/nsdi12/technical-sessions/presentation/singla},
  publisher     = {USENIX Association},
  month         = apr
}

@manual{nvidia-rail-optimize,
  title        = {Rail Optimized Topology Validation},
  author       = {{NVIDIA Corporation}},
  organization = {NVIDIA Networking},
  address      = {Santa Clara, CA},
  year         = {2025},
  month        = {Feb},
  note         = {Part of the ibdiagnet InfiniBand Fabric Diagnostic Tool User Manual; describes cabling validation and compute-fabric alignment in DGX SuperPOD rail‑optimized fabrics},
  url          = {https://docs.nvidia.com/networking/display/ibdiagnetusermanualv221/Rail+Optimized+Topology+Validation}
}

@misc{nvidia-pxn,
  author       = {{NVIDIA Corporation}},
  title        = {Doubling all‑to‑all Performance with NCCL 2.12: Introducing PXN (PCI X NVLink)},
  howpublished = {NVIDIA Developer Blog},
  month        = feb,
  year         = 2022,
  note         = {Describes PXN, which enables GPU‐to‐NIC communication via NVLink to optimize rail‐aligned collective performance},
  url          = {https://developer.nvidia.com/blog/doubling-all2all-performance-with-nvidia-collective-communication-library-2-12/}
}

@inproceedings {taccl,
author = {Aashaka Shah and Vijay Chidambaram and Meghan Cowan and Saeed Maleki and Madan Musuvathi and Todd Mytkowicz and Jacob Nelson and Olli Saarikivi and Rachee Singh},
title = {{TACCL}: Guiding Collective Algorithm Synthesis using Communication Sketches},
booktitle = {20th USENIX Symposium on Networked Systems Design and Implementation (NSDI 23)},
year = {2023},
isbn = {978-1-939133-33-5},
address = {Boston, MA},
pages = {593--612},
url = {https://www.usenix.org/conference/nsdi23/presentation/shah},
publisher = {USENIX Association},
month = apr
}

@misc{neye,
  author       = {{nEye Systems}},
  title        = {{nEye}: Dismantling Network Walls to Build a Sustainable {AI} Future},
  howpublished = {\url{https://www.neye.ai/}},
  year         = {2025},
  note         = {Optical circuit switch platform for AI datacenter networking. Accessed: 2025-02-02}
}

@misc{semianalysis2025xaicolossus,
  author       = {Ontiveros, Jeremie Eliahou and Patel, Dylan and Zhou, Wei},
  title        = {{xAI's Colossus 2 - First Gigawatt Datacenter In The World, Unique RL Methodology, Capital Raise}},
  howpublished = {SemiAnalysis},
  month        = sep,
  year         = {2025},
  url          = {https://newsletter.semianalysis.com/p/xais-colossus-2-first-gigawatt-datacenter},
  note         = {Accessed: 2026-01-23}
}

@inproceedings{morphlux,
author = {Vijaya Kumar, Abhishek and Ding, Eric and Devraj, Arjun and Bunandar, Darius and Singh, Rachee},
title = {Reconfigurable Torus Fabrics for Multi-tenant ML},
year = {2026},
isbn = {9798400723599},
publisher = {Association for Computing Machinery},
address = {New York, NY, USA},
url = {https://doi.org/10.1145/3779212.3790238},
doi = {10.1145/3779212.3790238},
booktitle = {Proceedings of the 31st ACM International Conference on Architectural Support for Programming Languages and Operating Systems, Volume 2},
pages = {1547–1565},
numpages = {19},
location = {USA},
series = {ASPLOS '26}
}

@inproceedings{astrasim,
  author    = {Saeed Rashidi and Srinivas Sridharan and Sudarshan Srinivasan and Tushar Krishna},
  title     = {{ASTRA-SIM: Enabling SW/HW Co-Design Exploration for Distributed DL Training Platforms}},
  booktitle = {{IEEE} International Symposium on Performance Analysis of Systems and Software, {ISPASS} 2020, Boston, MA, USA, August 22-26, 2020},
  publisher = {{IEEE}},
  year      = {2020},
}

@article{chakra,
  author    = {Srinivas Sridharan and Taekyung Heo and Louis Feng and Zhaodong Wang and Matt Bergeron and Wenyin Fu and Shengbao Zheng and Brian Coutinho and Saeed Rashidi and Changhai Man and Tushar Krishna},
  title     = {{Chakra: Advancing Performance Benchmarking and Co-design using Standardized Execution Traces}},
  journal   = {arXiv preprint arXiv:2305.14516},
  year      = {2023},
}

@misc{800g_tx,
  author       = {{FS.com Inc.}},
  title        = {{NVIDIA/Mellanox MMA4Z00-NS Optical Transceiver Module}},
  year         = {2025},
  howpublished = {\url{https://www.fs.com/products/229253.html}},
  note         = {Product page, Accessed: 2026-02-06}
}

@misc{800g_switch_power,
  author       = {{NVIDIA Corporation}},
  title        = {{NVIDIA Q32xx and Q34xx XDR 800Gb/s InfiniBand Switch Systems User Manual}},
  year         = {2024},
  howpublished = {\url{https://docs.nvidia.com/networking/display/xdrswitcheshwum/specifications}},
  note         = {Accessed: 2026-02-06}
}

@misc{800g_switch_price,
  author       = {{NADDOD}},
  title        = {{NVIDIA Quantum-X800 XDR InfiniBand Switch, Q3400-RA}},
  year         = {2025},
  howpublished = {\url{https://www.naddod.com/products/nvidia-networking/102612}},
  note         = {Reseller price listing, Accessed: 2026-02-06}
}

@misc{OCS_6000N,
  author       = {{Polatis}},
  title        = {{Polatis Series 6000n Optical Switch Datasheet}},
  year         = {2023},
  howpublished = {\url{https://www.redhelix.com/wp-content/uploads/2023/11/Polatis_6000n_Data_Sheet-rhl.pdf}},
  note         = {Datasheet, Accessed: 2026-02-06}
}

@misc{nvidia_mlnx_en,
  author       = {{NVIDIA Corporation}},
  title        = {{NVIDIA Ethernet Driver for Linux (mlnx\_en)}},
  howpublished = {\url{https://network.nvidia.com/products/ethernet-drivers/linux/mlnx_en/}},
  year         = {2025},
  note         = {Accessed: 2026-02-06}
}

@misc{nvidia_connectx6dx_firmware,
  author       = {{NVIDIA Corporation}},
  title        = {{ConnectX-6 Dx Firmware Download}},
  howpublished = {\url{https://network.nvidia.com/support/firmware/connectx6dx/}},
  year         = {2025},
  note         = {Accessed: 2026-02-06}
}

@misc{1600g_tx,
  author       = {{NADDOD}},
  title        = {{NVIDIA/Mellanox MMS4C11-XM-RHS Compatible 1.6T 2xDR4 OSFP Silicon Photonics Optical Transceiver Module}},
  year         = {2025},
  howpublished = {\url{https://www.naddod.com/products/103418.html}},
  note         = {Product page, Accessed: 2026-06-19}
}

@inproceedings{patel2024splitwise,
  title        = {{Splitwise}: Efficient Generative {LLM} Inference Using Phase Splitting},
  author       = {Patel, Pratyush and Choukse, Esha and Zhang, Chaojie and Shah, Aashaka and Goiri, {\'I}{\~n}igo and Maleki, Saeed and Bianchini, Ricardo},
  booktitle    = {2024 ACM/IEEE 51st Annual International Symposium on Computer Architecture (ISCA)},
  year         = {2024}
}

@inproceedings{zhong2024distserve,
  title        = {{DistServe}: Disaggregating Prefill and Decoding for Goodput-optimized Large Language Model Serving},
  author       = {Zhong, Yinmin and Liu, Shengyu and Chen, Junda and Hu, Jianbo and Zhu, Yibo and Liu, Xuanzhe and Jin, Xin and Zhang, Hao},
  booktitle    = {18th USENIX Symposium on Operating Systems Design and Implementation (OSDI 24)},
  year         = {2024}
}
\appendix
\clearpage
\section*{Appendix}
Appendices are supporting material that has not been peer-reviewed.
\section{Characteristics of different parallelism strategies}

\label{sec:charac}

ML systems leverage multiple, co-existing parallelisms (Table~\ref{tab:parallel_strategy}), with each parallelism type shown in Table~\ref{tab:paralllelism}. These parallelisms include data parallelism (DP and FSDP), pipeline parallelism (PP), tensor parallelism (TP, often with sequence parallelism or SP), context parallelism (CP), and expert parallelism (EP). Each parallelism incurs communication that differs in: data volume, start time, frequency, and communication pattern.

\begin{table}[h]
    \centering
    \footnotesize
    \begin{tabular}{ccc}
        \hline
        \textbf{Model size} & \textbf{Compute ($N$ GPUs)} & \textbf{Practices}\\
        \hline
        Small ($< 10$B) & $N \leq 8$ & TP or DP \\
        Large ($> 10$B) & $8 < N \leq 512$ & \footnotesize{TP \& PP, TP \& DP, or DP} \\
        Large ($> 10$B) & $512 < N \leq 1024$ & DP \& PP, or DP \& TP \\
        Large ($> 10$B) & $N > 1024$ & TP, DP \& PP \\
        \hline
    \end{tabular}
    \caption{Rule-of-thumb LLM parallelism strategies \cite{ultrascale_playbook}.}
    \label{tab:parallel_strategy}

\end{table}

\rev{
\begin{table*}[h]
    \centering
    \footnotesize
    \begin{tabular}{ccccc}
    \hline
    \textbf{Parallelism}    & \textbf{Memory reduction} & \textbf{Compute reduction} & \textbf{Communication type and frequency} & \textbf{Symmetrical?}\\
    \hline
    \hline
    DP        & \makecell{gbs/dp} & \makecell{gbs/dp} & \makecell{bwd AR per layer/per model} & yes \\
    \hline
    FSDP                     & \makecell{gbs/dp, params/dp} & \makecell{gbs/dp} & \makecell{fwd AG, bwd RS per layer/model} & yes\\
    \hline
    TP      & \makecell{params/tp, grads/tp, optims/tp} & \makecell{params/tp} & \makecell{fwd bwd AR per operator} & yes\\
    \hline
    TP \& SP     & \makecell{params/tp, grads/tp, optims/tp, activs/tp} & \makecell{params/tp, activs/tp} & \makecell{fwd bwd AG\&RS per operator} & yes\\
    \hline
    CP     & \makecell{kv\_cache/cp, seq/cp} &  \makecell{seq/cp} & \makecell{fwd AG bwd RS per layer} & yes\\
    \hline
    PP   & \makecell{params/pp, grads/pp, optims/pp, activs/pp} &  \makecell{params/pp}  & \makecell{fwd bwd \sendrecv per microbatch} & no\\
    \hline
    EP      & \makecell{experts/ep} &  \makecell{experts/ep} & \makecell{fwd bwd \alltoall per layer} & yes\\
    \hline
    \end{tabular}
    \caption{Characteristics of different parallelism strategies \cite{liang2024torchtitan}. 
        gbs: global batch size.
        dp: data parallel degree.
        seq: sequence length.
        fwd: forward pass.
        bwd: backward pass.
        AR: \allreduce.
        AG: \allgather.
        RS: \reducescatter.
        params: model parameter size.
        grads: gradients size.
        optims: optimizer states.
        activs: activation states.
    }
    \label{tab:paralllelism}
\end{table*}
}

\section{Additional Reconfiguration Window Study at Frontier Scale}
\label{sec:window-batch}

\rev{
\myparab{Setup.}
The frontier-scale characterization in Figure~\ref{fig:window}(d) and
Figure~\ref{fig:sim_util_window} uses a large global batch (15{,}360). Since
frontier model training often uses a varying global batch size in different training phases~\cite{liu2024deepseek}, we repeat the identical strong-scaling sweep
(GPT-style~\cite{man2025stage}, PP=4, TP=32, sequence length 4096, 96 layers, B200,
ASTRA-sim~\cite{astrasim}) at a $5\times$ smaller global batch of 3{,}072, sweeping
GPU count from 2{,}048 to 16{,}384. We use the same analytical pipeline as in \S~\ref{sec:trace}.}

\begin{table}[h]
    \centering
    \footnotesize
    \begin{tabular}{ccccccc}
        \hline
        \textbf{DP} & \textbf{GPUs} & \textbf{per-rank} & \textbf{mean} & \textbf{p5} & \textbf{p50} & \textbf{p95} \\
                    &               & \textbf{batch}    & \textbf{(ms)} & \textbf{(ms)} & \textbf{(ms)} & \textbf{(ms)} \\
        \hline
        $16$  & $2{,}048$  & $192$ & $136.0$ & $9.1$ & $37.9$ & $811.2$ \\
        $32$  & $4{,}096$  & $96$  & $69.1$  & $4.0$ & $19.3$ & $414.1$ \\
        $64$  & $8{,}192$  & $48$  & $35.7$  & $1.5$ & $9.9$  & $215.5$ \\
        $128$ & $16{,}384$ & $24$  & $19.2$  & $0.8$ & $5.2$  & $116.2$ \\
        \hline
    \end{tabular}
    \caption{\rev{Window size statistics (mean / p5 / p50 / p95, pooled over all
    rails) for a GPT-style workload under DP strong scaling
    (PP=4, TP=32, global batch size 3{,}072, sequence length 4096, 96 layers) running
    on B200.}}
    \label{tab:window3072}
\end{table}

\rev{
\myparab{Windows shrink with the batch.}
Table~\ref{tab:window3072} reports the window distribution at each scale.
Reducing the global batch also reduces the window sizes relative to the 15{,}360 case: the median (p50) window at 2{,}048 GPUs
falls from 187\,ms to 37.9\,ms, and from 23.9\,ms to 5.2\,ms at 16{,}384 GPUs. The distribution stays
wide---the shortest windows (p5) range from $9.1$\,ms at 2{,}048 GPUs down to $0.8$\,ms at 16{,}384 GPUs, while the longest (p95) remain in the hundreds of ms.}

\begin{figure}[t]
    \centering
    \includegraphics[width=\linewidth]{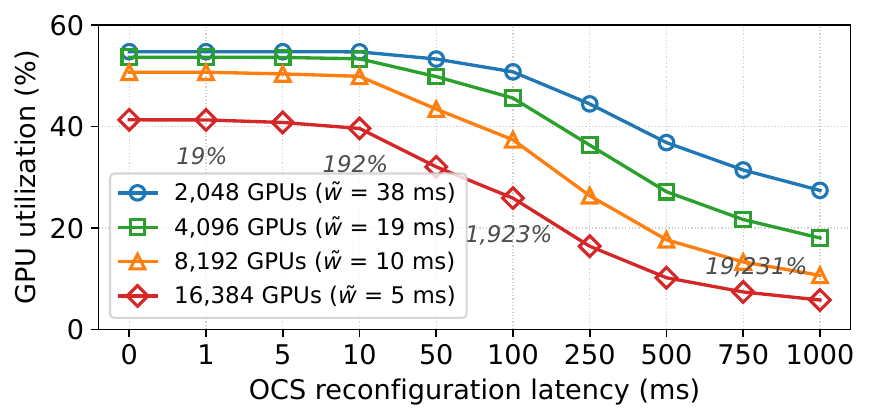}
    \caption{\rev{GPU utilization vs.\ OCS reconfiguration latency for a GPT-style
    workload on B200 at a smaller global batch of 3{,}072 (the same setting as in
    Table~\ref{tab:window3072}). $\tilde{w}$ is a workload's median window. 
    Percentages in the plot show the reconfiguration latency
    as a fraction of the median window at the largest (16{,}384-GPU) scale.}}
    \label{fig:util_window_3072}
\end{figure}

\rev{
\myparab{Utilization vs.\ reconfiguration latency.}
Figure~\ref{fig:util_window_3072} sweeps the OCS reconfiguration latency
$T_{\mathit{reconfig}}$ and reports the GPU utilization. Utilization is flat while $T_{\mathit{reconfig}}$ stays below most of the windows
($T_{\mathit{reconfig}} \le 10$\,ms) and then degrades if the latency is large. At 16{,}384 GPUs a 100\,ms reconfiguration reduces utilization
from 41.3\% to 25.9\%, whereas a sub-10\,ms OCS largely maintains it
(39.6\%). The results are consistent with the results for larger global batch size in \S~\ref{sec:trace},
showing that with fast (sub-10\,ms) OCS
switching, reconfiguration is hidden
inside the natural inter-phase windows, even when a smaller batch makes those
windows shorter.}

\section{\sysname Pseudocode}
\label{sec:code}

We provide complete pseudocode for the shim layer (Algorithm~\ref{alg:comm-pattern}).

\begin{algorithm*}[t]
    \caption{Communication Pattern Scheduling with Topology Reconfiguration}
    \label{alg:comm-pattern}
    \footnotesize
    \begin{algorithmic}[1]
    
    \State \textbf{Data:} \_phase\_table
    \State \quad $(start\_gid, start\_comm\_idx, end\_gid, end\_comm\_idx)$
    
    \State \textbf{Per-group state:}
    \State \quad $idx \gets 0$
    \State \quad $comm\_group \gets$ assigned\_group
    \State \quad $asym\_way \gets$ way\_in\_asymmetrical\_parallelism

    \State \textbf{Global state:}
    \State \quad $comm\_stage \gets 0$
    \State \quad $mode \in \{\textsc{provisioning}, \textsc{default}\}$
    
    \Function{phase\_change\_before}{}
        \State \Return $phase\_table[comm\_stage].start\_gid = comm\_group.id$ \textbf{and} $idx = phase\_table[comm\_stage].start\_comm\_idx$
    \EndFunction
    
    \Function{phase\_change\_after}{}
        \State \Return $phase\_table[comm\_stage].end\_gid = comm\_group.id$ \textbf{and} $idx = phase\_table[comm\_stage].end\_comm\_idx$
    \EndFunction
    
    \Function{get\_next\_comm}{}
        \If{\Call{phase\_change\_after}{}}
            \State \Return $(phase\_table[comm\_stage + 1].start\_gid,\; phase\_table[comm\_stage + 1].start\_comm\_idx)$
        \Else
            \State \Return $(comm\_group.id,\; idx + 1)$
        \EndIf
    \EndFunction

    \Procedure{pre\_comm}{$comm\_op$}
        \If{$comm\_op$ is scale\_up \textbf{or} management}
            \State{select scale-up or CPU front-end network}
            \State \Return
        \EndIf
        
        \State{wait till topology is free}
        \State $shift \gets \Call{phase\_change\_before}{}$ 
        \If{$mode = \textsc{default}$}
            \If{$shift$ \textbf{or} $comm\_op$ is asymm.}
                \State \Call{topo\_write}{$comm\_group.id,\; idx,\; asym\_way$}
            \EndIf
        \EndIf
    
        \If{$shift$}
            \State $comm\_stage \gets comm\_stage + 1$
            \State set topology busy
        \EndIf
    
        \State $idx \gets idx + 1$
        \State{select GPU backend network}
        \State register \textsc{post\_comm} callback 
    \EndProcedure
    
    \Procedure{post\_comm}{$comm\_group,\; comm\_op$}
        \State $shift \gets \Call{phase\_change\_after}{}$ 
        \If{$mode = \textsc{provisioning}$}
            \If{$shift$ \textbf{or} $comm\_op$ is asymm.}
                \State $(n\_group\_id,\; n\_idx) \gets \Call{get\_next\_comm}{}$
                \State \Call{topo\_write}{$n\_group\_id,\; n\_idx,\; asym\_way$}
            \EndIf
        \EndIf
    
        \If{$shift$}
            \State set topology free
        \EndIf
    \EndProcedure

    \end{algorithmic}
    \end{algorithm*}

\section{Rail Topologies for Expert-Parallel \alltoall}
\label{sec:alltoall-topo}

\rev{\myparab{Setup.}
A rail connects the GPUs of the same local rank across scale-up domains, so a
domain with $G$ GPUs has $G$ rails, each driven by its own OCS. An EP group of
size $E$ is spread over $m = E/G$ domains (assuming overlapping EP with TP), and its dispatch and combine \alltoall
must move data between every pair of these $m$ domains.
The default \sysname connects all domains with the same ring topology across rails.
On one ring, two domains can be up to $\lfloor m/2 \rfloor$ hops apart, so their traffic
is forwarded through intermediate domains, wasting bandwidth (a bandwidth tax).}

\rev{\myparab{Diverging rings across rails.}
To handle the large fan-out of \alltoall, \sysname instead gives each rail a
\emph{different} ring, so different rails connect different pairs of domains. This
trades scale-out hops for forwarding inside a domain, which is cheap because the
scale-up links have high bandwidth. Each ring connects $m$ pairs directly (every domain to
its two ring neighbors), so if a new rail's ring avoids the pairs already covered,
it adds $m$ new direct pairs and gives each domain two more direct neighbors. There
are $m(m-1)/2$ unique pairs in all, so $G$ rails connect at most $\min(Gm,\, m(m-1)/2)$ of
them directly, and the fraction of pairs reachable in a single scale-out hop is
\[
f(G, m) \;\le\; \min\!\left(1,\ \frac{Gm}{m(m-1)/2}\right) \;=\; \min\!\left(1,\ \frac{2G}{m-1}\right).
\]
Equality holds when the rings do not overlap. As long as $G \ge \lceil (m-1)/2 \rceil$ or $m \le 2G+1$ (
$E \le G(2G+1)$), every pair is directly connected ($f=1$) through \sysname with diverged rail topologies. With an NVL72 domain
($G=72$), for instance, this holds for EP groups of up to $G(2G+1) = 10{,}440$
GPUs, well above typical sizes ($E \le 128$).
}

\rev{\myparab{Ring assignment.}
A ring is an ordering $\pi$ of the domains, and the pairs it connects are the
neighbors in that ordering, $E(\pi) = \{\{\pi_k, \pi_{(k+1) \% m}\} : 0 \le k < m\}$.
Let $w(e)$ be the number of rails that already connect pair $e$.
Algorithm~\ref{alg:diverge} iteratively determines the rail ordering using a greedy method: for each rail it picks
the ring that adds the most new pairs, $u(\pi) = |\{e \in E(\pi) : w(e) = 0\}|$.
If there are ties, we select the ring whose pairs have been used the least so far
($\min \max_{e \in E(\pi)} w(e)$), which spreads any repeats evenly. 
}

\begin{algorithm}[t]
    \caption{Diverged ring assignment per EP group}
    \label{alg:diverge}
    \footnotesize
    \begin{algorithmic}[1]
    \Require domains $D$ ($|D| = m$), rails $G$, candidate orderings $C$
    \State $w(e) \gets 0$ for every pair $e \in \binom{D}{2}$ \Comment{\# rails wiring $e$}
    \For{$r = 1 \to G$}
        \ForAll{$\pi \in C$}
            \State $u(\pi) \gets |\{e \in E(\pi) : w(e) = 0\}|$ \Comment{new pairs}
        \EndFor
        \State $\pi_r \gets \displaystyle\arg\max_{\pi \in C}\ \big(\, u(\pi),\ -\!\!\max_{e \in E(\pi)} w(e) \,\big)$ \Comment{most new; tie: least reuse}
        \State $w(e) \gets w(e) + 1\ \ \forall e \in E(\pi_r)$
    \EndFor
    \State \Return $(\pi_1, \dots, \pi_G)$ \Comment{ring for each rail}
    \end{algorithmic}
\end{algorithm}

\rev{\myparab{Routing.}
\sysname control router sends each \sendrecv pair in \alltoall over a rail that connects its two domains
directly, splitting the load across all available rails. If there is an uncovered pair (large $m$, small $G$), 
the router selects the shortest multi-hop path instead, prioritizing forwarding in scale-up domains over scale-out. Since almost all pairs are direct, this
extra forwarding is negligible (\S\ref{sec:sim-ep}). The diverged topologies across rails are set up
before the dispatch \alltoall, overlapping with the expert-router computation
(\S\ref{sec:protocol}).}

\rev{\myparab{Simulating multi-hop forwarding.}
Our reconfigurable network backend (\S\ref{sec:sim}) models the forwarding cost of
these routes directly. 
We model store-and-forward instead of cut-through, so a chunk is fully received at each intermediate domain before being re-sent, and an $h$-hop route pays the per-link
latency plus the serialization delay (chunk size divided by link bandwidth) once
per hop. A multi-hop flow may also queue behind other flows at each hop. A
multi-hop route thus costs roughly $h\times$ the latency and link bandwidth of a
direct one, which is the penalty measured in \S\ref{sec:sim-ep}.}

\end{document}